\documentclass[10pt,twocolumn]{article}
\usepackage{overcite}
\usepackage{color}
\usepackage{graphicx}
\usepackage{graphics}
\usepackage{amsmath}
\usepackage{amssymb}

\usepackage{lettrine}

\title{
\textbf{Information-based fitness and the emergence of criticality in living systems}
}

\author {Jorge Hidalgo $^{1\dag}$, Jacopo Grilli, $^{2\dag}$, Samir Suweis
  $^{2}$, \\ Miguel A. Mu\~noz $^{1\ast}$,
  Jayanth R. Banavar $^{3}$, Amos Maritan $^{2\ast}$\\
  \\
  \footnotesize{$^{1}$ Departamento de
    Electromagnetismo y F{\'\i}sica de la Materia and}\\
 \footnotesize{ Instituto Carlos I
    de F{\'\i}sica Te\'orica y Computacional.}\\
  \footnotesize{Universidad de Granada.  E-18071, Granada, Spain.}\\
  \footnotesize{$^{2}$
    Dipartimento di Fisica `G. Galilei' \& CNISM, INFN.}\\
  \footnotesize{Universit\`a di Padova, Via Marzolo 8, 35131 Padova, Italy.}\\
  \footnotesize{$^{3}$ Department of Physics, University of Maryland,
  College Park, MD 20742, USA.}\\
  \footnotesize{$^{\dag}$ These authors contributed equally to this work.}\\
  \footnotesize{$^{\ast}$ To whom correspondence should be addressed: mamunoz@onsager.ugr.es, amos.maritan@pd.infn.it. }  
}
\date{June 2014}
\newcommand{\ang}[1]{\langle #1\rangle}

\DeclareMathOperator*{\argmin}{arg\,min}
\DeclareMathOperator*{\argmax}{arg\,max}

\begin{document}

\twocolumn[
  \begin{@twocolumnfalse} 


\maketitle


\begin{abstract} 
Empirical evidence suggesting that living systems might operate in the vicinity of critical points, at the borderline between order and disorder, has proliferated in recent years, with
examples ranging from spontaneous brain activity to flock dynamics.  However, a well-founded theory for understanding how and why interacting living systems could dynamically tune themselves to be
poised in the vicinity of a critical point is lacking.
Here we employ tools from statistical mechanics and information theory to show that complex adaptive or evolutionary systems can be much more efficient in coping with diverse heterogeneous
environmental conditions when operating at criticality.  Analytical as well as computational evolutionary and adaptive models vividly illustrate that a community of such systems dynamically self-tunes
close to a critical state as the complexity of the environment increases while they remain non-critical for simple and predictable environments.  A more robust convergence to criticality emerges in
co-evolutionary and co-adaptive set-ups in which individuals aim to represent other agents in the community with fidelity, thereby creating a collective critical ensemble and providing the best
possible trade-off between accuracy and flexibility.  Our approach provides a parsimonious and general mechanism for the emergence of critical-like behavior in living systems needing to cope with
complex environments or trying to efficiently coordinate themselves as an ensemble.
 \end{abstract}

\newpage

 \end{@twocolumnfalse}
]

\clearpage

Physical systems undergo phase transitions from ordered to disordered states on changing control parameters \cite{stanley1987,Binney}. Critical points,
with all their remarkable properties
\cite{stanley1987,Binney}, are only observed upon parameter fine tuning.  This is in sharp contrast to the ubiquity of critical-like behavior in complex living matter. Indeed, empirical evidence has
proliferated that living systems might operate at criticality~\cite{Bialek2011} --i.e. at the borderline between order and disorder-- with examples ranging from spontaneous brain
behavior~\cite{Beggs2003}, gene expression patterns~\cite{Kauffman08}, cell growth~\cite{Kaneko}, morphogenesis \cite{Bialek2014}, bacterial clustering~\cite{chen2012scale}, and flock
dynamics~\cite{Cavagna2012}. Even if none of these examples is fully conclusive and even if the meaning of ``criticality'' varies across these works, the criticality hypothesis --as a general
strategy for the organization of living matter-- is a tantalizing idea worthy of further investigation.

Here we present a framework for understanding how self-tuning to criticality can arise in living systems.  Unlike models of self-organized criticality in which some inanimate systems are found to
become critical in a mechanistic way \cite{Jensen}, our focus here is on general adaptive or evolutionary mechanisms, specific to biological systems. We suggest that the drive to criticality arises
from functional advantages of being poised in the vicinity of a critical point.

But why is a living system fitter when it is critical? Living systems need to perceive and respond to environmental cues and to interact with other similar entities. Indeed, biological systems
constantly try to encapsulate the essential features of the huge variety of detailed information from their surrounding complex and changing environment into manageable internal representations, and
they use these to base their actions and responses.  The successful construction of these representations, which extract, summarize, and integrate relevant information \cite{Adami11}, provides a
crucial competitive advantage, which can eventually make the difference between survival and extinction.  We suggest here that criticality is an optimal strategy to effectively represent the
intrinsically complex and variable external world in a parsimonious manner.  This is in line with the hypothesis that living systems benefit from having attributes akin to criticality --either
statistical or dynamical \cite{Bialek2011}-- such as a large repertoire of dynamical responses, optimal transmission and storage of information, and exquisite sensitivity to environmental changes
\cite{Binney,Kauffman08,Chialvo10,Beggs2008,Kinouchi-Copelli,mora2010maximum,Plenz2013}.

As conjectured long ago, the capability to perform complex computations, which turns out to be the fingerprint of living systems, is enhanced in ``machines'' operating near a critical point
\cite{Langton1990,Ber-Nat,Kauffman93}, i.e. at the border between two distinct phases: a disordered phase, in which perturbations and noise propagate unboundedly --thereby corrupting information
transmission and storage -- and an ordered phase where changes are rapidly erased, hindering flexibility and plasticity. The marginal, critical, situation provides a delicate compromise between these
two impractical tendencies, an excellent trade-off between reproducibility and flexibility \cite{Chialvo10,Plenz2013,Beggs2008} and, on larger time scales, between robustness and evolvability
\cite{Wagner}.  A specific example of this general framework are genetic regulatory networks \cite{Kauffman93,Gros08}. Cells ranging from those in complex organisms to single-celled microbes such as
bacteria respond to signals in the environment by modifying the expression of their genes. Any given genetic regulatory network, formed by the genes (nodes) and their interactions (edges)
\cite{DeJong2002}-- can be tightly controlled to robustly converge to a fixed almost-deterministic attractor --i.e. a fixed ``phenotype''-- or it can be configured to be highly sensitive to tiny
fluctuations in input signals, leading to many different attractors, i.e. to large phenotypic variability \cite{Huang2010}. These two situations correspond to the ordered and disordered phases
respectively.  The optimal way for genetic regulatory networks to reconcile controllability and sensitivity to environmental cues is to operate somewhere in between the two limiting and impractical
limits alluded to above~\cite{Kauffman93} as has been confirmed in different experimental set-ups \cite{Kauffman08,Bialek2014,GRN-experimental2}. Still, it is not clear how such tuning to criticality
comes about.

Our goal here is to exploit general ideas from statistical mechanics and information theory to construct a quantitative framework showing that self-tuning to criticality is a convenient strategy
adopted by living systems to effectively cope with the intrinsically complex external world in an efficient manner thereby providing an excellent compromise between accuracy and flexibility.  In order
to provide some further intuition, we employ genetic regulatory networks as a convenient guiding example, but one could equally well consider neural networks, models for the immune response, groups of
animals exhibiting collective behavior, etc., with each specific realization requiring a more detailed modeling of its special attributes.

We uncover co-evolutionary and co-adaptive mechanisms by which communities of living systems, even in the absence of other forms of environmental complexity, converge to be almost critical in
the process of understanding each other and creating a ``collective entity''.  The main result is that criticality is an evolutionary/adaptive stable solution reached by living systems in their
striving to cope with complex heterogeneous environments or when trying to efficiently coordinate themselves as an ensemble.

\begin{figure*}[htb]
\centering
\includegraphics[width=\textwidth]{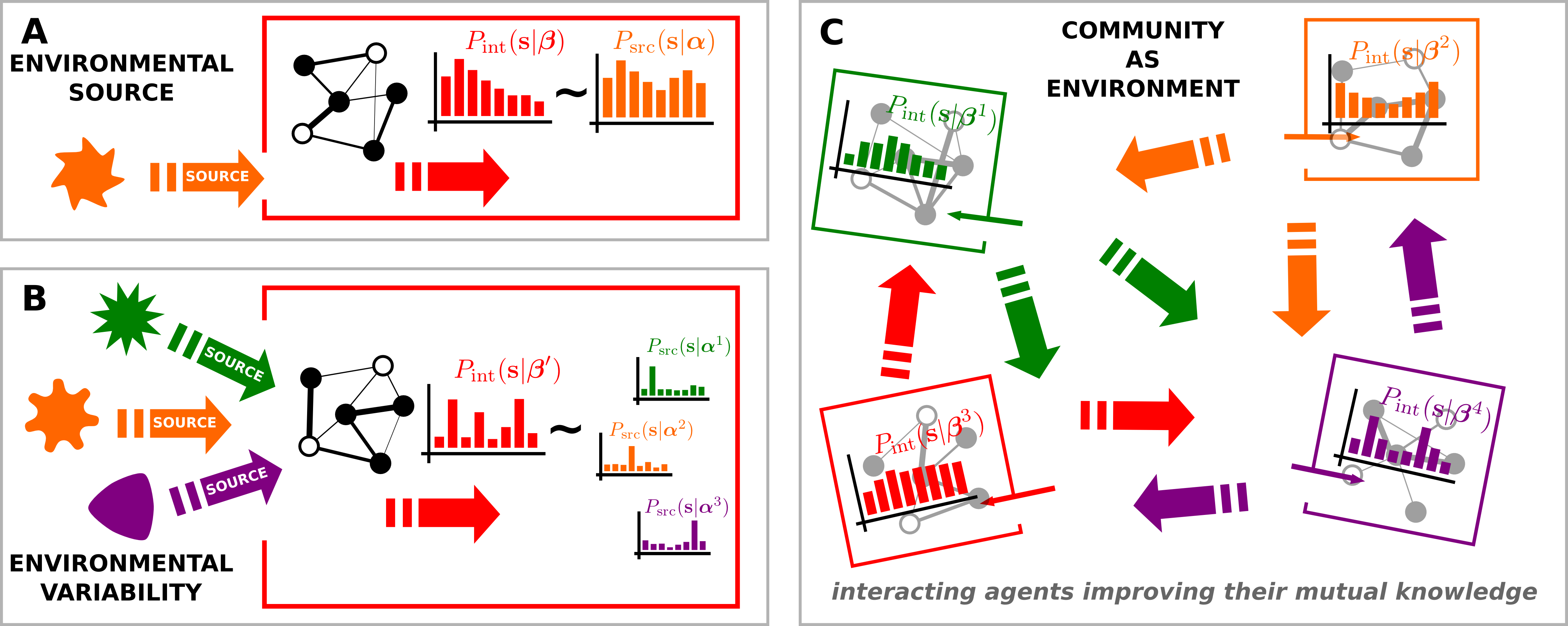}
  \caption{ \footnotesize {\bf Living systems coping with the environment.}  Panel A illustrates a {\it living system} responding to an {\it environmental source} (e.g. a bacteria responding to some
external conditions such as the presence/absence of some nutrients, pH concentration, or temperature). A given source, labeled by the set of parameters $\boldsymbol{\alpha}$, can only be
probabilistically gauged by the system.
    $P_{\mathrm{src}}(\mathbf{s}|\boldsymbol{\alpha})$ is the most accurate representation that the system can potentially generate in terms of the Boolean variables (or bits) $\mathbf{s}$.  However,
such a representation might not be accessible to the system by merely changing its internal-state parameters, $\boldsymbol{\beta}$, and the actual internal state,
$P_{\mathrm{int}}(\mathbf{s}|\boldsymbol{\beta})$, (e.g. the probability of a gene expression pattern) is usually an imperfect proxy for $P_{\mathrm{src}}(\mathbf{s}|\boldsymbol{\alpha})$.  The
optimal choice of parameters $\boldsymbol{\beta}$ --aiming at capturing the most relevant features of the environment-- is obtained by minimizing the Kullback-Leibler divergence of
$P_{\mathrm{int}}(\mathbf{s}|\boldsymbol{\beta})$ from $P_{\mathrm{src}}(\mathbf{s}|\boldsymbol{\alpha})$. In genetic networks, changing internal parameters is equivalent to changing the interactions
between the different (Boolean) variables (nodes of the networks in the figure). Panel B shows a more complex scenario, where the system has to cope with multiple and diverse sources. The internal
state has to be able to accommodate each of them.  In panel C, the environment is not imposed {\it ad hoc} but instead, it is composed of other individuals, and every agent needs to cope with
(``understand'') the states of the others.  Each agent evolves similarly to the others in the community, trying to exhibit the same kind of state, generating in this way a self-organized environment. 
In the case of sufficiently heterogeneous externally imposed sources as well as in the self-organized case, we find that evolutionary/adaptive dynamics drive the systems to operate close to
criticality.}
\label{fig:model} \end{figure*}

\section{Results}

\subsection{Mathematical framework}

The external environment in which living systems operate is highly variable, largely unpredictable, and describable in terms of probability distribution functions. Living systems need to modify their
internal state to cope with external conditions and they do so in a probabilistic manner.  To be specific, but without loss of generality, we represent an environmental cue ``perceived'' and
processed by a living system as a string of $N$ (binary) variables, $\mathbf{s}=(s_1, s_2,\ \dots\, s_N)$.  A specific environmental source is modeled by the probability distribution
$P_{\mathrm{src}}$ with which it produces each of the $2^N$ possible states. For concreteness, this distribution is assumed to depend on a set of parameters, $\boldsymbol{\alpha}=(\alpha_1,\ \alpha_2,
\dots )$, accounting for environmental variability.  We turn now to an individual living system or ``agent'', which seeks to adapt itself to cope with the perceived stimuli/signals emanating from a
given environmental source. This is accomplished by changing its internal state, encapsulated in a second probability distribution function, $P_{\mathrm{int}}$, specified by a different
--smaller in principle-- parameter set $\boldsymbol{\beta}=(\beta_1,\ \beta_2, \dots)$ aimed at capturing the essential features of $P_{\mathrm{src}}$ in the most efficient --though in general
imperfect-- way (see Figure 1).  Henceforth we will denote the external source and its internal representation by $P_{\mathrm{src}}(\mathbf{s}|\boldsymbol{\alpha})$ and
$P_{\mathrm{int}}(\mathbf{s}|\boldsymbol{\beta})$ respectively.

In our guiding example, the external cues could be, for instance, the environmental (temperature, pH,...) conditions, which are variable and can only be probabilistically gauged by a
cell/bacterium.  The binary vector $\mathbf{s}=(s_1, s_2,\ \dots\, s_N)$ can be thought of as the on/off state of the different $N$ genes in its (Boolean) genetic regulatory network
\cite{Kauffman93,Gros08,DeJong2002}. In this way, $P_{\mathrm{src}}(\mathbf{s}|\boldsymbol{\alpha})$ can be interpreted as the probability that the most convenient state aimed at by the system to cope
with a given environmental condition is $\mathbf{s}$, while $P_{\mathrm{int}}(\mathbf{s}|\boldsymbol{\beta})$ is the actual probability for the genetic-network state (attractor) of a given individual
--with its limitations-- to be $\mathbf{s}$.  Without loss of generality, we consider that there is (at least) one control parameter, say $\beta_1$, such that --other parameters being fixed-- it
determines in which phase the network is operating.

Our thesis is that the capacity of living systems to tune their internal states to efficiently cope with variable external conditions provides them with a strong competitive advantage. Thus, the
internal state $P_{\mathrm{int}}(\mathbf{s}|\boldsymbol{\beta})$ should resemble as closely as possible the one most in accord with the environmental signal
$P_{\mathrm{src}}(\mathbf{s}|\boldsymbol{\alpha})$; in other words, one seeks the distribution that the system should express in order to best respond to the external conditions.  Information theory
provides us with a robust measure of the ``closeness'' between the aimed (source) and the actual (internal) probability distribution functions.  Indeed, the Kullback-Leibler (KL) divergence
\cite{Cover-Thomas}, $D(\boldsymbol{\alpha}|\boldsymbol{\beta})$, quantifies the information loss when the internal state is used to approximate the source (see Materials and Methods, MM). The KL
divergence is asymmetric in the two involved probability distributions, it is never negative, and it vanishes if and only if the two distributions are identical (see SI Appendix, sec. S2).  Minimizing
the KL divergence with respect to the internal-state parameters, $\boldsymbol{\beta}$, generates the optimal, though in general imperfect, internal state aimed at representing or coping-with a given
source (see Fig.~\ref{fig:model}A).

More generally, in an ever-changing world, the requirement for an individual is not just to reproduce a single source with utmost fidelity but rather to be able to successfully cope with a group of
highly diverse sources (see Fig.~\ref{fig:model}B). A particularly interesting example of this would comprise a community of similar individuals which together strive to establish some kind of a
common collective language (see Fig.~\ref{fig:model}C). In any of these complex situations, our working hypothesis is that an individual has a larger ``fitness'' when a characteristic measure,
e.g.  the mean, of its KL divergences from the set of diverse sources is small, i.e. fit agents are those whose internal states are close to those required by existing external conditions.

As an illustrative example, consider two individual agents A and B --the source for A is B and vice versa-- each of them with its own probabilistic gene network. The relative fitnesses of A and
B are determined by 
how well the set of cues (described by the probability distribution $P_\mathrm{src}$) of one organism is captured by the other with minimum information loss, and vice versa
(for utter simplicity, we could assume that the distributions associated with A and B
correspond to equilibrium distributions of an Ising model
\cite{stanley1987,Binney} at similar inverse temperatures $\beta_A$ and $\beta_B$).  If $\beta_A=\beta_B$, the two distributions would be identical and the KL divergence
would vanish. However, this is not a stable solution. Indeed, if the two parameters are not identical but close, the 
difference between their respective KL divergences from each to the other is (see MM):
\begin{equation}
D(\beta_A+\delta\beta|\beta_A) - D(\beta_A|\beta_A+\delta\beta) \simeq   \frac{1}{6}\nabla \chi(\beta_A) \delta\beta^3
\end{equation}
where $\chi$ is the generalized susceptibility also known as ``Fisher information'' (defined in MM).  This implies that the individual whose parameters correspond to the state with larger $\chi$ has a
smaller KL divergence and is thus fitter. But it is well-known that $\chi$ peaks at the critical point, and thus our key finding is that, for a family of individuals with similar parameters, the
fittest possible agent sits exactly at criticality, and it is best able to encapsulate a wide variety of distributions.  As we illustrate in what follows with a number of examples, the optimal
encoding parameters of stable solutions lie always around the peak of the generalized susceptibility $\chi$ which is the region of maximal variability, where different complex sources can be best
accounted for through small parameter changes (see MM).  This is in line with the recent finding --based on concepts of information geometry-- that many more distinguishable outputs can be reproduced
by models poised at the peak of $\chi$, i.e. at criticality \cite{Marsili2011}.

\subsection{Computational experiments}

We have developed diverse computational evolutionary and adaptive models exploiting the ideas above. The dynamical rules employed in these models are \emph{not} meant to, necessarily, mimic the actual
dynamics of living systems, rather they are efficient ways to optimize fitness. In the evolutionary models, inspired by the genetic algorithm \cite{GA1,Gros08}, a community of $M$ individuals --each
one characterized by its own set of internal parameters $\boldsymbol{\beta}$-- evolves in time through the processes of death, birth, and mutation (see MM).  Individuals with larger fitness, i.e. with
a smaller mean KL divergence from the rest of sources, have a larger probability to produce an offspring, which --apart from small random mutations-- inherits its parameters from its ancestor. On the
other hand, agents with low fitness are more likely to die and be removed from the community. In the adaptive models, individuals can change their internal parameters if the attempted variation
implies an increase of their corresponding fitnesses (see MM). These evolutionary/adaptive rules result in the ensemble of agents converging to a steady state distribution, which we aim at
characterizing.  We obtain similar results in two families of models, which differ in the way in which the environment is treated. In the first, the environment is self-generated by a community of
co-evolving/co-adapting individuals, while, in the second, the variable external world is defined {\it ad hoc}.

\subsection*{Co-evolutionary model} 
The environment perceived by each individual consists of the other $M-1$ systems in the community, which it aims at ``understanding'' and coping with.  In the simplest computational implementation of
this idea (see MM), a pair of individual agents is randomly selected from the community at each time step and each of these two individuals constitutes the environmental source for the other. Given
that the KL divergence is not symmetric (see MM), one of the two agents has a larger fitness and thus a greater probability of generating progeny, while the less fit system is more likely to die. This
corresponds to a fitness function of agent $i$ which is a decreasing function of the KL divergence from the other.  In this case, as illustrated in Fig.~\ref{fig:coevol} (and in Supplementary Videos 1
and 2), the co-evolution of $M=100$ agents --which on their turn are sources-- leads to a very robust evolutionarily-stable steady-state distribution.  Indeed, the different left panels show that for
three substantially different initial parameter distributions (very broad, and localized in the ordered and in the disordered phases, respectively) the community co-evolves in time to a unique
localized steady state distribution, which turns out to be peaked at the critical point (i.e. where the Fisher information peaks, see Fig.~\ref{fig:coevol} right panel and SI Appendix, sec. S4).  This
conclusion is robust against model details and computational implementations: the solution peaked at criticality is an evolutionary stable attractor of the dynamics. The same conclusions hold for an
analogous “Co-adaptive model” in which the systems adapt rather than dying and replicating (see SI Appendix, sec. S6).

\begin{figure*}[htb]
  \centering
\includegraphics[width=\textwidth]{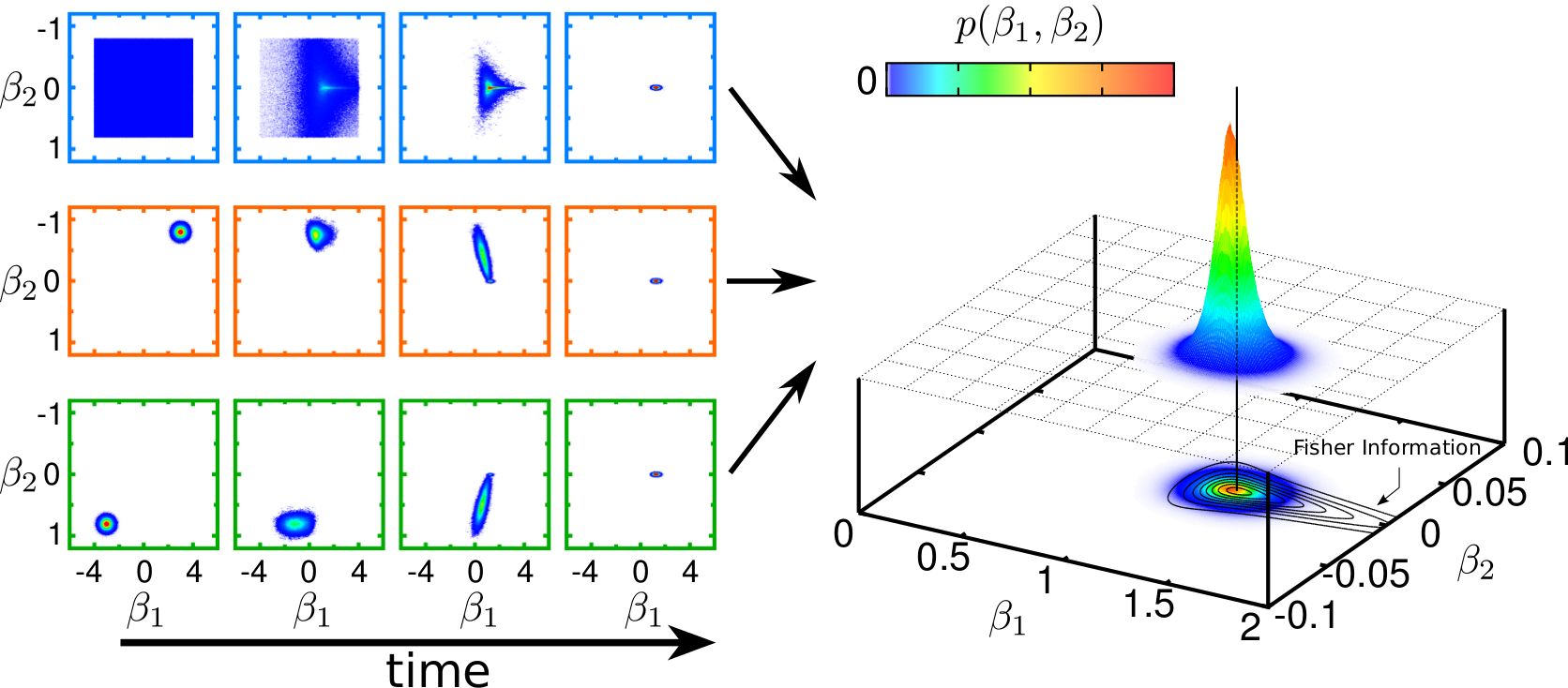}
  \caption{ \footnotesize {\bf Co-evolutionary model leads self-consistently to criticality:} A community of $M$ living systems (or agents) evolves according to a genetic algorithm dynamics
\cite{GA1}. Each agent
$i$ ($i=1,...,M$) is characterized by a 2-parameter ($\beta^i_1, \beta^i_2$) internal state distribution $P_{\mathrm{int}}(\mathbf{s}|\beta_1^i,\beta_2^i)$ and the rest of the community acts as the
external environment it has to cope with, i.e. the agents try to ``understand'' each other.  At each simulation step, two individuals are randomly chosen and their respective relative fitnesses are
computed in terms of the Kullback-Leibler divergence from each other's internal state probability distribution.  One of the two agents is removed from the community with a probability that is smaller
for the fitter agent; the winner produces an offspring which (except for small variations/mutations) inherits its parameters. (Left) These co-evolutionary rules drive the community very close to
a unique
localized steady state. As shown (Right), this is localized precisely at the critical point, i.e. where the generalized susceptibility or Fisher information of the internal state
distribution exhibits a sharp peak (as shown by the contour plots and heat maps). The internal state distributions are parameterized as $P_{\mathrm{int}}(\mathbf{s}|\beta_1,\beta_2)\propto
\exp{\{\beta_1 \frac{N}{2} (\sum_{k=1}^N \frac{s_k}{N})^2 + \beta_2 \sum_{k=1}^N s_k\}}$ representing a global (all-to-all) coupling of the internal nodes (see MM).  Much more complex probability
distributions in which all units are not coupled to all other units --i.e. more complex networked topologies-- are discussed in the SI Appendix, sec. S4.}
 \label{fig:coevol} 
 \end{figure*}

\subsection*{Evolutionary model}
\begin{figure}
\centering
\includegraphics[width=1.\columnwidth]{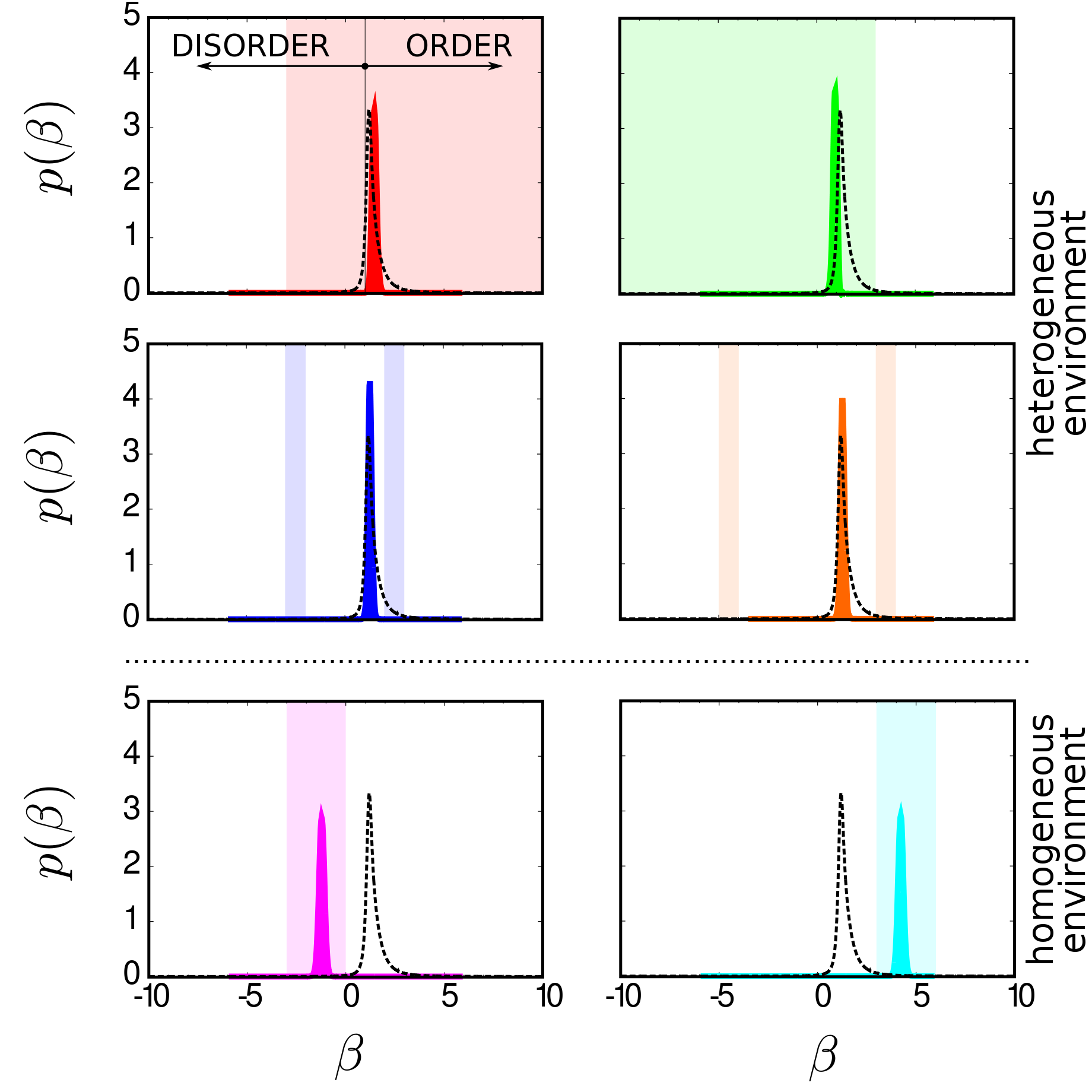}
  \caption{ \footnotesize {\bf Evolutionary model leading to near to criticality in complex environments}.  A community of $M$ agents undergoes a genetic algorithm dynamics \cite{GA1}.  Each agent is
simultaneously
exposed to diverse stimuli $\mathbf{s}$ provided by $S$ different sources, each one characterized by a probability $P_{\mathrm{src}}(\mathbf{s}|\alpha^u)$ with $u=1,\dots, S$, fully specified by
parameters $\alpha^u$.  At each time step, $S$ sources are randomly drawn with probability $\rho_{\mathrm{src}}(\alpha^u)$ (in this case a uniform distribution with support in the colored region). 
Each agent $i$ ($i=1,...,M$) has an internal state $P_{\mathrm{int}}(\mathbf{s}|\beta^i)$ aimed at representing --or coping with-- the environment. Agents' fitness increases as the mean
Kullback-Leibler divergence from the set of sources to which they are exposed decreases. The higher the fitness of an individual, the lower its probability of dying.  An agent that is killed is
replaced by a new individual with a parameter $\beta$ inherited from one of the other agents (and, with some probability, a small variation/mutation). The community dynamically evolves and eventually
reaches a steady state distribution of parameters, $p(\beta)$. The six panels in the figure correspond to different supports (colored regions) for uniform source distributions,
$\rho_{\mathrm{src}}(\alpha^u)$. The dashed line is the generalized susceptibility (Fisher information) of the internal probability distribution, which exhibits a peak at the critical point separating
an ordered from a disordered phase.  Heterogeneous source pools (Top and Middle) lead to distributions peaked at criticality, whereas for homogeneous sources (Bottom), the communities
are not critical but specialized. Stimuli distributions are parametrized in a rather simplistic way as
$P_{\mathrm{src}}(\mathbf{s}|\alpha^u)\propto\exp\{\alpha^u\frac{N}{2}(\sum_{k=1}^N\frac{s_k}{N})^2\}$, while internal states are identical but replacing $\alpha^u$ by $\beta^i$ (see MM).  In the
guiding example of genetic regulatory networks, this example corresponds to an extremely simple fully-connected network in which the state of each gene is equally determined by all the other genes
and, hence, the probability of a given state depends only on the total number of on/off genes, controlled by a single parameter.}
\label{fig:evol}
\end{figure}

An ensemble of $M$ agents are exposed at each particular time to a heterogeneous complex environment consisting of $S$ independent environmental sources, each one with a different $P_{\mathrm{src}}$
and thus parametrized by diverse $\boldsymbol{\alpha}$s (see Fig.~\ref{fig:evol}).  The set of $S$ sources is randomly extracted from a broadly distributed pool of possible sources occurring with
different probabilities, $\rho_{\mathrm{src}}(\boldsymbol{\alpha})$. The fitness of an individual with parameters $\boldsymbol{\beta}$ with respect to any given environment is taken to be a decreasing
function of the average KL divergence from the diverse external stimuli: $\displaystyle d(\rho_{\mathrm{src}} | \boldsymbol{\beta}):= \int d \boldsymbol{\alpha}
\rho_{\mathrm{src}}(\boldsymbol{\alpha}) D(\boldsymbol{\alpha}|\boldsymbol{\beta}) \ $. In the special case of just one internal-state parameter, $\beta_1$, we find that upon iterating the genetic
algorithm (see MM), the distribution evolves towards a steady-state stable distribution.  Computer simulations show that in the case of very homogeneous environments, occurring when all sources in the
pool are similar --$\rho_{\mathrm{src}}(\boldsymbol{\alpha})$ sufficiently narrow-- the optimal $\boldsymbol{\beta}$ strongly depends on the specific sources, resulting in detail-specific internal
states (see bottom panels in Fig.~\ref{fig:evol}).  On the other hand, if the external world is sufficiently heterogeneous (see SI Appendix, sec. S5) the optimal internal state becomes peaked near the
critical point (see top panels in Fig.~\ref{fig:evol} and Supplementary Video 3 illustrating the evolution of agents towards the vicinity of the critical point).
We note that the approach to criticality is less precise in this model than in the co-evolutionary one, in which the environment changes progressively as the agents co-evolve, allowing the system to
systematically approach the critical point with precision. Similar conclusions hold for an analogous ``adaptive model'' (see SI Appendix, sec. S6).  Finally, one might wonder whether the resulting
closeness to criticality in these models is not just a byproduct of the environment itself being critical in some way. In fact, it has been recently shown that complex environments, when hidden
variables are averaged out, can be effectively described by the Zipf's law \cite{Schwab}, a signature of criticality \cite{Bialek2011}.  This observation applies to some of the heterogeneous
environments analyzed here which, indeed, turn out to be Zipfian; however, as shown in the SI Appendix, sec. S5, there are simple yet heterogeneous environments, which are not Zipfian but nevertheless
result in the same behavior.

\subsection{Analytical results for the dynamical models}

A generic probability distribution can be rewritten to parallel the standard notation in statistical physics, $P(\mathbf{s}|\boldsymbol{\gamma})={\exp \big(-H(\mathbf{s}|\boldsymbol{\gamma}) \big)} /
{Z(\boldsymbol{\gamma})} \ $, where the factor $Z(\boldsymbol{\gamma})$ is fixed through normalization. The function $H$ can be generically written as $H(\mathbf{s}|\boldsymbol{\gamma})=\sum_{\mu}
\gamma_\mu \phi^{\mu}(\mathbf{s}) \ $, where $\phi^{\mu}(\mathbf{s})$ are suitable functions (``observables'') of the variables $\mathbf{s}$.  For a specific set of parameters $\boldsymbol{\alpha}$
characterizing an environmental source, the best possible internal state --minimizing the KL divergence-- can be shown to obey
$\ang{\phi_{\mathrm{int}}^\mu}_{\boldsymbol{\alpha}} = \ang{\phi_{\mathrm{int}}^\mu}_{\boldsymbol{\beta}} \ , $
where the index $\mu$ runs over the whole set of parameters and $\ang{\phi_{\mathrm{int}}^\mu}_{\boldsymbol{\alpha}} := \sum_{\mathbf{s}}\phi_{\mathrm{int}}^\mu(\mathbf{s})
P_{\mathrm{int}}(\mathbf{s}|\boldsymbol{\alpha})$ and $\ang{\phi_{\mathrm{int}}^\mu}_{\boldsymbol{\beta}} := \sum_{\mathbf{s}}\phi_{\mathrm{int}}^\mu(\mathbf{s})
P_{\mathrm{src}}(\mathbf{s}|\boldsymbol{\beta})$. This result implies that the optimal internal state is the one which best reproduces the lowest moments of the original source distribution it seeks
to cope with (the number of moments coinciding with --or being limited by-- the number of free parameters). By evaluating the second derivatives (Hessian matrix) it is easy to verify that, if a
solution exists, it actually corresponds to a minimum of the KL divergence (see SI Appendix, sec. S3).

To proceed further, we need to compute the internal-state distribution in the presence of diverse sources distributed with $\rho_{\mathrm{src}}(\boldsymbol{\alpha})$.  In this case, we compute the
value of $\boldsymbol{\beta}$ which minimizes the \textit{average KL divergence to the sources} $\boldsymbol{\alpha}$ as written above (an alternative possibility --which is discussed in the SI
Appendix, sec. S3-- is to identify the optimal $\boldsymbol{\beta}$ for each specific source and then average over the source distribution), leading to the condition:
 $ \displaystyle \ang{\phi_{\mathrm{int}}^\mu}_{\boldsymbol{\beta}} = \int d \boldsymbol{\alpha} \ \rho_{\mathrm{src}}(\boldsymbol{\alpha}) \ \ang{\phi_{\mathrm{int}}^\mu}_{\boldsymbol{\alpha}}$.
 We consider the simple example in which both the sources and the system are characterized by a single parameter and, assuming that a phase transition occurs at some parameter value $\alpha=\alpha_c$,
i.e. $\big<\phi\big>_{\alpha}$ has a sigmoid shape (which becomes steeper as $N$ increases) with an inflection point at $\alpha=\alpha_c$ (our analysis can be extended to more general cases where
there is no built-in phase transition in the source distributions but they are merely sufficiently heterogeneous). The two plateaus of the sigmoid function correspond to the so-called
\textit{disordered} and \textit{ordered} phases, respectively. When $\rho_{\mathrm{src}}(\alpha)$ has support on both sides of the sigmoid function, i.e. when it is ``heterogeneous'', by solving the
equation for the optimal $\beta$ it is obvious that the moment to be reproduced lies somewhere in between the two asymptotic values of the sigmoid with the values of $\beta$ for which intermediate
moments are concentrated near the inflection or critical point, $\alpha_c$. Indeed, as $\chi =-\frac{d <\phi>_{\beta}}{ d \beta}$, the critical region, where the generalized susceptibility $\chi$ has
a peak, is the region of maximal variability in which different complex sources can be best accounted for through small parameter changes, in agreement with the finding that many more distinguishable
outputs can be reproduced by models poised close to criticality \cite{Marsili2011}.

\section{Discussion and conclusions}
 
Under the mild assumption that living systems need to construct good though approximate internal representations of the outer complex world and that such representations are encoded in terms of
probability distributions, we have shown --by employing concepts from statistical mechanics and information theory-- that the encoding probability distributions do necessarily lie where the {\it
generalized susceptibility} or {\it Fisher information} exhibits a peak \cite{Cover-Thomas}, i.e.  in the vicinity of a critical point, providing the best possible compromise to accommodate both
regular and noisy signals.

In the presence of broadly different ever-changing heterogeneous environments, computational evolutionary and adaptive models vividly illustrate how a collection of living systems eventually cluster
near the critical state.  A more accurate convergence to criticality is found in a co-evolutionary/co-adaptive set-up in which individuals evolve/adapt to represent with fidelity other agents in the
community, thereby creating a collective ``language'', which turns out to be critical.

These ideas apply straightforwardly to genetic and neural networks --where they could contribute to a better understanding of why neural activity seems to be tuned to criticality-- but have a broader
range of implications for general complex adaptive systems \cite{Gros08}.  For example, our framework could be applicable to some bacterial communities for which a huge phenotypic (internal state)
variability has been empirically observed \cite{Leibler2005}. Such a large phenotypic diversification can be seen as a form of ``bet hedging'', an adaptive survival strategy analogous to stock-market
portfolio management \cite{Wolf1}, which turns out to be a straightforward consequence of individuals in the community being critical. Usually, from this point of view, generic networks diversify
their ``assets'' among multiple phenotypes to minimize the long-term risk of extinction and maximize the long-term expected growth rate in the presence of environmental uncertainty \cite{Wolf1}. 
Similar bet-hedging strategies have been detected in viral populations and could be explained as a consequence of their respective communities having converged to a critical state, maximizing the
hedging effect.  Similarly, criticality has been recently shown to emerge through adaptive information processing in machine learning, where networks are trained to produce a desired output from a
given input in a noisy environment; when tasks of very different complexity need to be simultaneously learned, networks adapt to a critical state to enhance their performance \cite{Rohlf12}.  In
summary, criticality in some living systems could result from the interplay between their need for producing accurate representations of the world, their need to cope with many widely diverse
environmental conditions, and their well-honed ability to react to external changes in an efficient way. Evolution and adaptation might drive living systems to criticality in response to this smart
cartography.


\section{Materials}

\textbf{Kullback-Leibler divergence}.
Given two probability distributions $P(\mathbf{s})$ and $Q(\mathbf{s})$ for variables $\mathbf{s}$, the
Kullback-Leibler (KL) divergence of $Q(\mathbf{s})$ from $P(\mathbf{s})$,
\begin{equation}
D(P|Q) := \sum_{\mathbf{s}} P(\mathbf{s})
\log \Bigl( \frac{P(\mathbf{s})}{Q(\mathbf{s})} \Bigr) \ ,
\label{eq:KL}
\end{equation}
quantifies the loss of information when $Q(\mathbf{s})$ is used to approximate $P(\mathbf{s})$~\cite{Cover-Thomas}. Indeed, in the large $T$ limit, the probability $\mathcal{L}$ that the model
$Q(\mathbf{s})$ generates a sequence of $T$ observations compatible with $P(\mathbf{s})$ can be computed as $ \mathcal{L} \sim \exp \Bigl( -T D(P|Q) \Bigr) $ up to leading order (see SI Appendix, sec.
 S2).  Therefore, maximizing the likelihood of a trial probability distribution function $Q$ is equivalent to minimizing its KL divergence from the original one, $P$. In the Results section we
use the
notation $D(\boldsymbol{\alpha}|\boldsymbol{\beta})$ when $P(\mathbf{s})=P_{\mathrm{src}}(\mathbf{s}|\boldsymbol{\alpha})$ and $Q(\mathbf{s})=P_{\mathrm{int}}(\mathbf{s}|\boldsymbol{\beta})$.

\textbf{Fisher information and criticality}.
Given a probability distribution $P(\mathbf{s}|\boldsymbol{\gamma})$
--where $\boldsymbol{\gamma}$ can stand either for  $\boldsymbol{\alpha}$ or
$\boldsymbol{\beta}$
-- the Fisher Information  is defined as
\begin{equation}
\chi^{\mu\nu}(\boldsymbol{\gamma}):=\displaystyle \left\langle
\frac{\partial \log P(\cdot|\boldsymbol{\gamma})}{\partial
\gamma_\mu}\frac{\partial \log P(\cdot|\boldsymbol{\gamma})}{\partial
\gamma_\nu} \right\rangle_{\boldsymbol{\gamma}},
\end{equation}
where $\mu$ and $\nu$ are parameter labels and the average $\langle \cdot \rangle_{\boldsymbol{\gamma}}$ is performed with respect to $P(\cdot|\boldsymbol{\gamma})$. It measures the {\it amount of
information} encoded in the states $\mathbf{s}$ about the parameters $\boldsymbol{\gamma}$ \cite{Cover-Thomas}.  This follows from the Cram\'er-Rao inequality, which states that the error made when we
estimate $\boldsymbol{\gamma}$ from one state $\mathbf{s}$ is, on average, greater (or at least equal) than the inverse of the Fisher information \cite{Cover-Thomas}.  In particular, if $\chi$ happens
to diverge at some point, it is possible to specify the associated parameters with maximal precision \cite{Marsili2011}.
 With the parametrization used in the main text, the Fisher information  is the {\it generalized susceptibility} in the statistical mechanics terminology and measures the response of the system to
parameter variations:
$\chi^{\mu\nu}(\boldsymbol{\gamma}) = -\frac{\partial \langle \phi^\mu
\rangle_{\boldsymbol{\gamma}}}{\partial \gamma_\nu}
= \langle\phi^\mu \phi^\nu\rangle_{\boldsymbol{\gamma}} -
\langle\phi^\mu \rangle_{\boldsymbol{\gamma}}
\langle\phi^\nu\rangle_{\boldsymbol{\gamma}}  \ ,$
and is well-known to peak at critical points \cite{stanley1987,Binney}.

\textbf{Co-evolutionary model}. The $k$-th agent of the community is described by a probability distribution $P_{\mathrm{int}}(\mathbf{s}|\boldsymbol{\beta}^k)\propto
\exp\{-H_{\mathrm{int}}(\mathbf{s}|\boldsymbol{\beta}^k)\}$, with $ H_{\mathrm{int}}(\mathbf{s}|\boldsymbol{\beta}^k) = \sum_{\mu}^{I} \beta_\mu^k \phi^\mu_{\mathrm{int}}(\mathbf{s})$, depending on
parameters $\boldsymbol{\beta}^k$.  Starting with an ensemble of $M$ agents whose internal parameters are extracted from an arbitrary distribution, $p(\boldsymbol{\beta})$, two individuals, $i$ and
$j$, are randomly selected at each time step. Their relative fitnesses $f^{(j)}_i$ and $f^{(i)}_j$ are computed as
$
f^{(j)}_i=1-
{D(\boldsymbol{\beta}^j|\boldsymbol{\beta}^i)}/[{D(\boldsymbol{\beta}^j|\boldsymbol{\beta}^i)+D(\boldsymbol{\beta}^i|\boldsymbol{\beta}^j)]}   \ ,
$
and similarly for $f_j^{(i)}$ (as the KL divergence is not symmetric, $f_i^{(j)} \neq f_j^{(i)}$ unless $\boldsymbol{\beta}^i=\boldsymbol{\beta}^j$).  One of the two individuals --selected with
probability equal to its relative fitness-- creates an offspring, while the other one is removed from the community.  The offspring inherits its parameters from its ancestor (with prob. $1-\nu$) or
mutates with a probability $\nu$, modifying its parameters from $\boldsymbol{\beta}$ to $\boldsymbol{\beta}\rightarrow \boldsymbol{\beta}+ \boldsymbol{\xi}$, where $\boldsymbol{\xi}$ is a multivariate
Gaussian random vector, with uncorrelated components, zero mean and deviation $\boldsymbol{\sigma}$. Time is updated to $t\rightarrow t + 1/M$, another couple of individuals $i'$ and $j'$ is picked,
and the
process is iterated.  Variants of this model are described in the SI Appendix, sec. S4.

\textbf{Evolutionary model}. A community of agents receiving external stimuli from an outer and heterogeneous environment is modeled as follows. Every specific environmental source corresponds to a
probability distribution $P_{\mathrm{src}}(\mathbf{s}|\boldsymbol{\alpha})\propto \exp\left(- H_{\mathrm{src}}(\mathbf{s}|\boldsymbol{\alpha})\right)$, with $
H_{\mathrm{src}}(\mathbf{s}|\boldsymbol{\alpha}) = \sum_{\mu}^{E} \alpha_\mu \phi^\mu_{\mathrm{src}}(\mathbf{s})$, where the parameters $\boldsymbol{\alpha}$ are drawn from the distribution
$\rho_{\mathrm{src}}(\boldsymbol{\alpha})$.  The $k$-th agent in the community constructs an internal representation of the observed source described by
$P_{\mathrm{int}}(\mathbf{s}|\boldsymbol{\beta}^k)\propto \exp\left(- H_{\mathrm{int}}(\mathbf{s}|\boldsymbol{\beta}^k)\right)$ with $ H_{\mathrm{int}}(\mathbf{s}|\boldsymbol{\beta}^k) =
\sum_{\mu}^{I} \beta_\mu^k \phi^\mu_{\mathrm{int}}(\mathbf{s}), $ with parameters $\boldsymbol{\beta}^k$.  We start with $M$ individuals each one equipped with some initial parameter set extracted
from some arbitrary
  distribution $p(\boldsymbol{\beta})$.  At every time step, we generate $S$ external sources,
  $\{\boldsymbol{\alpha}^u\}_{u=1,...,{S}}$, from the source-pool
  $\rho_{\mathrm{src}}(\boldsymbol{\alpha})$. Then we compute the average KL divergence of every individual's internal state distribution from the external sources
$ 
 d(\{\boldsymbol{\alpha}^u\}|\boldsymbol{\beta}^k) :=
\sum_{u=1}^{S} D(\boldsymbol{\alpha}^u|\boldsymbol{\beta}^k)/S
$
The $k$-th individual of the community is removed with a
  probability proportional to its average KL divergence (or any increasing function of it)
$
P_{\mathrm{kill}}(k) = {d(\{\boldsymbol{\alpha}^u\}|\boldsymbol{\beta}^k)}/{\sum_{l}d(\{\boldsymbol{\alpha}^u\}|\boldsymbol{\beta}^{l}) }
$
and it is replaced by an offspring of another individual randomly selected from the rest of the community.  The offspring inherits its parameters 
from the parent and time is updated as in the co-evolutionary model. 

\section{Acknowledgments}
We are indebted to T. Hoang, D. Pfaff, J. Uriagereka, S. Vassanelli, and M. Zamparo for useful discussions and to W. Bialek and two anonymous referees for many insightful
suggestions. A.M., J.G. and S.S. acknowledge Cariparo foundation for financial support. M.A.M. and J.H. acknowledge support from J. de Andalucia P09-FQM-4682 and the Spanish MINECO FIS2009-08451.


\newpage
\onecolumn

\setcounter{equation}{0} 
\setcounter{section}{0} 
\setcounter{figure}{0}
\setcounter{table}{0}
\setcounter{page}{1}

\begin{center}
\begin{LARGE}
\textbf{
Supplementary Information:\\
Information-based fitness and the emergence of criticality in living systems}
\end{LARGE}

\hspace*{1cm}

J. Hidalgo, J. Grilli, S. Suweis, M.A. Mu\~noz, J.R. Banavar, and A. Maritan
\end{center}

\vspace{0.5cm}

\parskip=2mm
\parindent=0mm

\renewcommand{\thesection}{S\arabic{section}}
\renewcommand{\figurename}{Supplementary Figure}
\renewcommand{\tablename}{Supplementary Table}
\renewcommand{\thefigure}{S\arabic{figure}}
\renewcommand{\thetable}{S\arabic{table}}
\renewcommand{\theequation}{S\arabic{equation}}
\renewcommand{\thepage}{S\arabic{page}} 
\newcommand{\parcial}[2]{\frac{\partial #1}{\partial #2}}

\makeatletter
\newenvironment{figurehere}{\def\@captype{figure}}{}
\makeatother

\newcommand{\DeclareCaption}[2]{\def#1{\def\@captype{#2}\caption}}

\DeclareCaption{\figcaption}{figure}
\DeclareCaption{\tabcaption}{table}

\section{Brief primer on critical phenomena}
\label{sec:criticality}
Critical phenomena are well understood in physical systems
\cite{Stanley1987-S, Binney-S}. The great lesson learned from statistical
physics is that, even though the elementary constituents can be as
simple as spheres or spins sitting on a lattice with pair-wise
interactions, criticality and scale invariance emerge as the
collective behavior of a many-body system with its characteristics
depending only on just a few essential attributes such as the
dimensionality of the system and symmetries of the
problem. Remarkably, this universality results in the critical
behavior of an Ising model with nearest neighbor interactions on a
cubic lattice being identical to that of a liquid-vapor system at its
critical point. Likewise, the critical behavior of a binary alloy that
is about to order is the very same as the two other cases. This is
because all these systems are three dimensional and have the same
``up-down'' symmetry. The underlying details: the fact that spins sit
on an idealized lattice, the chemistry of the liquid, or the atomic
interactions in an alloy are irrelevant in determining the critical
behavior.

A classical Ising spin can point up or down. An interaction between
neighboring spins results in favoring a parallel relative orientation
over an antiparallel one. This is captured through an interaction
energy that is lower and thus more favorable when neighboring spins
are parallel compared to when they are antiparallel. Note that our
description of the model favors parallel over antiparallel but up and
down are treated symmetrically. The ground state or the lowest energy
state of such an Ising system is one in which all spins are parallel
and, by necessity, are all up or all down. This choice between up and
down breaks the up-down symmetry spontaneously. The advantage of
lowering the energy through a mostly parallel alignment and thus
breaking the symmetry, favored at low temperatures, competes with the
tendency to increase the entropy at high temperatures through a
restoration of the symmetry and having roughly equal numbers of up and
down spins. While there are just two states with perfectly parallel
spins --all up or all down-- there are many states and therefore a
higher entropy when approximately half the spins are up and the other
half are down. The magnetization of such an Ising system is
proportional to a suitably normalized imbalance between the numbers of
up and down spins. It provides a measure of the ordering and is zero
at high temperatures and is $1$ at zero temperature. On lowering the
temperature from a very high value, the magnetization remains zero
until a critical temperature is reached at which point the
magnetization rises continuously and becomes non-zero. The critical
point is then a special temperature at which there is an onset of a
non-zero magnetization and the up-down symmetry is spontaneously
broken. There are two phases that emerge: at any temperature higher
than $T_c$, the magnetization is zero for an infinite sized system
whereas, below $T_c$, the magnetization is non-zero. The critical point,
which separates these two phases, is obtained by tuning the
temperature just right to its critical value. At a critical point,
there are domains of up and down spins of all sizes thoroughly
interspersed among one another. Scale invariance occurs because there
is no dominant size scale associated with these domains and power law
correlations between spin orientations are observed.

At the critical point, the system is exquisitely sensitive to external
perturbations of the right sort. For example, imposing a magnetic
field, whose tendency is to align the spins along the field direction,
does not have a significant effect at very low temperatures because
the spins are aligned parallel to each other and there is already a
non-zero magnetization. A measure of this sensitivity is provided by
the magnetic susceptibility, which is a measure of a differential
increase of magnetization due to an imposition of an infinitesimal
magnetic field. As argued before, the susceptibility is small at low
temperatures. It is also small at very high temperatures – thermal
effects, which favor high entropy, do not encourage much alignment of
the spins and imposing a tiny magnetic field does not lead to much
alignment. As a function of temperature, the susceptibility shows a
peak at the critical temperature. Remarkably, for an infinite system,
the susceptibility becomes infinitely large at its critical point. For
a finite size system, the position of the maximum of the
susceptibility provides an excellent measure of the location of the
critical point. We use a generalized susceptibility in the main text
as a diagnostic of critical behavior.

\section{Kullback-Leibler divergence, maximum likelihood, and Sanov's theorem}
\label{sec:KL}

Given two probability distributions $P(\mathbf{s})$ and
$Q(\mathbf{s})$ for the set of variables $\mathbf{s}$, the
Kullback-Leibler (KL) divergence of $Q(\mathbf{s})$ from
$P(\mathbf{s})$ is defined as
\begin{equation}
D(P(\cdot)|Q(\cdot)) := \sum_{\mathbf{s}} P(\mathbf{s})
\log \Bigl( \frac{P(\mathbf{s})}{Q(\mathbf{s})} \Bigr) \ ,
\label{eq:KL-S}
\end{equation}
and quantifies the loss of information when $Q(\mathbf{s})$ is
used to approximate $P(\mathbf{s})$~\cite{kullback1951-S,Cover-Thomas-S}.
The KL divergence is non-negative and it vanishes if and only if both
distributions are equal.  Observe also that the KL
divergence is not symmetric and therefore is not a properly-defined
``distance''.

The KL divergence can be understood in terms of the maximum likelihood
principle or of the Sanov's theorem \cite{Sanov-S,Cover-Thomas-S,Mezard-S}.
Consider a long sequence of empirical data consisting of $T$
independent measurements.  Let $C(\mathbf{s})$ be the number of times
a certain event $\mathbf{s}$ is repeated in the sequence. Suppose that
events are distributed as $P(\mathbf{s})$. In the large $T$ limit, the
frequencies $C(\mathbf{s})/T$ converge to $P(\mathbf{s})$, by the
Glivenko-Cantelli theorem ~\cite{Cover-Thomas-S}. A model for a finite
sampling of size $T$ can be represented as a probability distribution
$Q(\mathbf{s})$. The (multinomial) likelihood is defined as
\begin{equation}
\mathcal{L} = \frac{T!}{\prod_{\mathbf{s}} C(\mathbf{s})! }
\prod_{\mathbf{s}}  Q(\mathbf{s})^{C(\mathbf{s})} \ ,
\label{eq:MultLike}
\end{equation}
which is nothing but the probability that the model $Q(\mathbf{s})$
generates a sequence of $T$ observations compatible with $C(\mathbf{s})$.
The previous equation can be rewritten as
\begin{equation}
\mathcal{L} = T! ~ \exp{ \sum_{\mathbf{s}} \Bigl( C(\mathbf{s})
\log \big( Q(\mathbf{s}) \big) - \log \big( C(\mathbf{s})! \big) \Bigr)}
\ ,
\label{eq:MultLike2}
\end{equation}
which, in the large $T$ limit, using the approximations $T!\sim T^T$ and $ C(\mathbf{s}) \sim T
P(\mathbf{s})$ becomes
\begin{equation}
  \mathcal{L} \sim  \exp \Bigl(  -T D(P(\mathbf{\cdot})|Q(\mathbf{\cdot})) \Bigr)
\label{eq:MultLike3}
\end{equation}
up to leading order.  Therefore, maximizing the likelihood of a trial
probability distribution function $Q$ is equivalent to minimizing its
KL divergence with respect to the original one, $P$. This result is
also known as Sanov's theorem~\cite{Sanov-S} in the context of large
deviations theory.

\section{Representing the external world}
\label{sec:mapping}

An environmental cue (called a ``source'' from now on) can be modeled
by a set of $N$ binary variables $\mathbf{s}= (s_1, s_2,\ \dots\,
s_N)$, where $s_i= \pm 1, \forall i$. Sources $\mathbf{s}$ are
distributed with a suitable
$P_{\mathrm{src}}(\mathbf{s}|\boldsymbol{\alpha})$, which specifies
the statistical properties of sources and depends on some
``environmental'' parameters $\boldsymbol{\alpha}=(\alpha_1,\
\alpha_2, \dots, \alpha_{E})$. It can be interpreted as the
probability that the state $\mathbf{s}$ is the most adequate response
to the environment.

Without loss of generality, it is possible to write
\begin{equation}
  P_{\mathrm{src}}(\mathbf{s}|\boldsymbol{\alpha})=
  \frac{\exp \big(- H_{\mathrm{src}}(\mathbf{s}|\boldsymbol{\alpha}) \big)}{Z_{\mathrm{src}}(\boldsymbol{\alpha})} \ ,
\label{eq:Hsrc}
\end{equation}
which defines $H_{\mathrm{src}}$ up to a constant, independent of
$\mathbf{s}$, which can be set equal to zero. The factor
$Z_{\mathrm{src}}(\boldsymbol{\alpha})$ is defined by normalization
condition.  The quantity $H_{\mathrm{src}}$ can be expressed as
\begin{equation}
H_{\mathrm{src}}(\mathbf{s}|\boldsymbol{\alpha})=\sum_{\mu=1}^{E} \alpha_\mu \phi_{\mathrm{src}}^{\mu}(\mathbf{s}) \ ,
\label{eq:Hsrc2}
\end{equation}
where $E$ is the number of parameters specifying the source and
$\phi^{\mu}(\mathbf{s})$ are suitable functions (``observables'').
Equations \ref{eq:Hsrc} and \ref{eq:Hsrc2} are a convenient parametrization of the probability
distribution function, so that different values of $\boldsymbol{\alpha}$ specify distinct source distributions $P_{\mathrm{src}}$.

An individual agent is characterized by a probability
$P_{\mathrm{int}}(\mathbf{s}|\boldsymbol{\beta})$, depending on a set
of internal parameters $\boldsymbol{\beta}=(\beta_1,\ \beta_2, \dots,
\beta_{I})$, which represents the actual response to the environment.
In analogy with eq.~\ref{eq:Hsrc},
$P_{\mathrm{int}}(\mathbf{s}|\boldsymbol{\beta})$ can be written as
\begin{equation}
P_{\mathrm{int}}(\mathbf{s}|\boldsymbol{\beta})=\frac{\exp
\big(- H_{\mathrm{int}}(\mathbf{s}|\boldsymbol{\beta}) \big)}{Z_{\mathrm{int}}(\boldsymbol{\beta})} \ ,
\label{eq:Hmap}
\end{equation}
where
\begin{equation}
H_{\mathrm{int}}(\mathbf{s}|\boldsymbol{\beta})=\sum_{\mu=1}^{I} \beta_\mu \phi_{\mathrm{int}}^{\mu}(\mathbf{s}) 
\label{eq:Hmap2}
\end{equation}
where $I$ is
the number of ``internal'' parameters.

Information theory provides us with a natural tool to characterize an
optimal choice of $\boldsymbol{\beta}$ for a given
$\boldsymbol{\alpha}$.  In particular, the Kullback-Leibler (KL)
divergence (see section~\ref{sec:KL}) quantifies the amount of
information lost when a probability distribution is approximated by
another one.  Minimizing the Kullback-Leibler divergence between the
``source'' and ``internal'' distributions leads to the following
general relation between the source's parameters $\boldsymbol{\alpha}$
and the optimal choice for the internal parameter set \begin{equation}
  \boldsymbol{\beta}= \mathbf{L}(\boldsymbol{\alpha}) =
  \argmin_{\beta'} D(\boldsymbol{\alpha}|\boldsymbol{\beta}') \ ,
\label{eq:Lg}
\end{equation}
where $D(\boldsymbol{\alpha}|\boldsymbol{\beta})$ is the KL divergence
of equations ~\ref{eq:Hsrc} and ~\ref{eq:Hmap}.  The ability to cope
--even if in an optimal way-- with a single and precise environmental
source is not enough to cope with a complex and changing world.
Therefore, we need to introduce a last key ingredient: {\it
  environmental variability}.  In order to account for broadly diverse
and variable external sources, $\boldsymbol{\alpha}$, we introduce a
probability density of different parameter sets
$\rho_{\mathrm{src}}(\boldsymbol{\alpha})$, which describes the
variability of $\boldsymbol{\alpha}$ or the probability to encounter a
source with a given choice of parameters.


\subsection{Analytical results I: General theory}
\label{sec:analytic}

The KL divergence between the distributions characterized by generic
parameter sets $\boldsymbol{\alpha}$ and $\boldsymbol{\beta}$ respectively,
$D(\boldsymbol{\alpha}|\boldsymbol{\beta})=
D(P_{\mathrm{src}}(\cdot|\boldsymbol{\alpha})|P_{\mathrm{int}}(\cdot|\boldsymbol{\beta})) $,
can be easily written by using the generic parametrization of
equations~\ref{eq:Hsrc} and~\ref{eq:Hmap}:
\begin{equation}
D(\boldsymbol{\alpha}|\boldsymbol{\beta}) = \sum_{\mu=1}^{I} \beta_\mu
\ang{\phi_{\mathrm{int}}^\mu}_{\boldsymbol{\alpha}} + \log Z_{\mathrm{int}}(\boldsymbol{\beta}) - S_{\mathrm{src}}(\boldsymbol{\alpha}) \ ,
\label{eq:KLgen}
\end{equation}
with
\begin{eqnarray}
\displaystyle
\ang{\phi_{\mathrm{int}}^\mu}_{\boldsymbol{\alpha}} :=  \sum_{\mathbf{s}}
\phi_{\mathrm{int}}^\mu(\mathbf{s}) P_{\mathrm{src}}(\mathbf{s}|\boldsymbol{\alpha}) \\
\displaystyle
S_{\mathrm{src}}(\boldsymbol{\alpha}) := - \sum_{\mathbf{s}}
P_{\mathrm{src}}(\mathbf{s}|\boldsymbol{\alpha})
\log \Bigl(  P_{\mathrm{src}}(\mathbf{s}|\boldsymbol{\alpha})  \Bigr) \ ,
\label{eq:ave} 
\end{eqnarray}

where the last expression is the entropy of the distribution $ P_{\mathrm{src}}$.

To proceed further, there are two alternative ways to define the
optimal choice of parameter $\boldsymbol{\beta}$ given the distribution of
external parameters $\rho_{\mathrm{src}}(\boldsymbol{\alpha})$.

In the first one (to which we will refer as ``quenched'' choice), we
construct an internal response for every source, characterized by
$\boldsymbol{\alpha}$, by minimizing the KL divergence with respect to
$\boldsymbol{\beta}$, and then we compute the distribution of internal
states $\rho_{\mathrm{int}}(\boldsymbol{\beta})$. The optimal choice,
$\boldsymbol{\beta}^{\text{opt}}_{\text{q}}$, corresponds to the
maximum value of $\rho_{\mathrm{int}}$, i.e. the value of
$\boldsymbol{\beta}$ which is able to cope with the higher number of
sources.  Mathematically, it can be express as
\begin{equation}
 \boldsymbol{\beta}^{\text{opt}}_{\text{q}} =\argmax_{\boldsymbol{\beta}} \Bigl[ \int d \boldsymbol{\alpha} \delta \big( \boldsymbol{\beta} -
 \argmin_{\boldsymbol{\beta}'}
 D(\boldsymbol{\alpha}|\boldsymbol{\beta}') \big) \rho_{\mathrm{src}}(\boldsymbol{\alpha}) \Bigr] \ ,
\label{eq:optmapQue}
\end{equation}
Instead, in the second (``annealed'') case we compute the value of $\boldsymbol{\beta}$ which has a lower average KL divergence over the parameters $\boldsymbol{\alpha}$. 
Formally, it is written as
\begin{equation}  \boldsymbol{\beta}^{\text{opt}}_{\text{a}} =\argmin_{\boldsymbol{\beta}} \Bigl[ \int d \boldsymbol{\alpha}
  D(\boldsymbol{\alpha}|\boldsymbol{\beta}) \big) \rho_{\mathrm{src}}(\boldsymbol{\alpha}) \Bigr] \
\label{eq:optmapAn}
\end{equation}
and the system adopts the single value of $\boldsymbol{\beta}$,
$\boldsymbol{\beta}^{\text{opt}}_{\text{a}}$, that better describes
{\it on average} the varying environment. The annealed choice
represents the principal case studied in the main text of this work.


\subsubsection{Analytical results for the quenched choice}
\label{sec:quenched}

Given a set of parameters $\boldsymbol{\alpha}$ characterizing the source, the
optimal representation $\mathbf{L}(\boldsymbol{\alpha})$ is defined by
eq.~\ref{eq:Lg}.  By using eq.~\ref{eq:KLgen}, one readily obtains the equations for the stationary points
(extrema) of the KL divergence
\begin{equation}
0=\parcial{}{\beta_\mu} D(\boldsymbol{\alpha}|\boldsymbol{\beta}) =
\ang{\phi_{\mathrm{int}}^\mu}_{\boldsymbol{\alpha}} - \ang{\phi_{\mathrm{int}}^\mu}_{\boldsymbol{\beta}} \ ,
\label{eq:stat}
\end{equation}
with the index $\mu$ running from $1$ to $I$ 
and $\ang{\phi_{\mathrm{int}}^\mu}_{\boldsymbol{\beta}} :=  \sum_{\mathbf{s}}\phi_{\mathrm{int}}^\mu(\mathbf{s}) P_{\mathrm{int}}(\mathbf{s}|\boldsymbol{\beta})$.

To check whether extrema are minima or maxima, one needs to evaluate
the Hessian matrix at the stationary point,
\begin{equation}
  \frac{\partial^2}{\partial \beta_\mu \partial \beta_\nu}
  D(\boldsymbol{\alpha}|\boldsymbol{\beta}) = - \parcial{}{\beta_\nu}
  \ang{\phi_{\mathrm{int}}^\mu}_{\boldsymbol{\beta}} = \ang{\phi_{\mathrm{int}}^\mu
    \phi_{\mathrm{int}}^\nu}_{\boldsymbol{\beta}}
  - \ang{\phi_{\mathrm{int}}^\mu}_{\boldsymbol{\beta}}\ang{\phi_{\mathrm{int}}^\nu}_{\boldsymbol{\beta}}
  \equiv
  \ang{\ang{\phi_{\mathrm{int}}^\mu \phi_{\mathrm{int}}^\nu}}_{\boldsymbol{\beta}} \ ,
\label{eq:statHessian}
\end{equation}
which is a positive defined matrix (excluding the trivial case of a
factorized $P_{\mathrm{int}}(\mathbf{s}|\boldsymbol{\beta})$ for which
it vanishes). Therefore, if the solution of eq.~\ref{eq:stat} exists,
it corresponds to minimum of the KL
divergence. Equation~\ref{eq:stat}, which implicitly defines the
optimal internal parameters
$\boldsymbol{\beta}=\mathbf{L}(\boldsymbol{\alpha})$, has an intuitive
interpretation: {\it the minimum of the KL divergence is obtained when
  the first $I$ moments of the distribution
  $P_{\mathrm{int}}(\mathbf{s}|\boldsymbol{\beta})$ exactly match
  those of $P_{\mathrm{src}}(\mathbf{s}|\boldsymbol{\alpha})$}.

Therefore, the internal parameters are uniquely fixed via the optimal
solution of eq.~\ref{eq:stat} given the internal ones.  If we introduce
a distribution of external parameters
$\rho_{\mathrm{src}}(\boldsymbol{\alpha})$ it is straightforward to
obtain a distribution of internal parameters
$\rho_{\mathrm{int}}(\boldsymbol{\beta})$, defined in the following
way
\begin{equation}
  \displaystyle
  \rho_{\mathrm{int}}(\boldsymbol{\beta}) := \int  d \boldsymbol{\alpha} \
\rho_{\mathrm{src}}(\boldsymbol{\alpha})
\delta \big( \boldsymbol{\beta} - \mathbf{L}(\boldsymbol{\alpha}) \big)
 \ .
\label{eq:jacob0}
\end{equation}
One can show that this distribution is proportional to the {\it Fisher
  information} or {\it generalized susceptibility} of the internal
system,
\begin{equation}
\displaystyle
\chi_{\mathrm{int}}^{\mu\nu}(\boldsymbol{\beta})  := - \parcial{}{\beta_\mu} \ang{\phi_{\mathrm{int}}^\nu}_{\boldsymbol{\beta}}
= \ang{\ang{\phi_{\mathrm{int}}^\mu \phi_{\mathrm{int}}^\nu}}_{\boldsymbol{\beta}}  \ ,
\label{eq:suscep_mapappa}
\end{equation}
which diverges at the critical point in the limit $N\to\infty$. When
the external parameters are not centered about a critical point, the
distribution $\rho_{\mathrm{int}}(\boldsymbol{\beta})$ will peak at
the critical point at large but finite $N$.

\subsubsection{Analytical results for the annealed choice}
\label{sec:annealed}
Here we analyze the model defined via eq.~\ref{eq:optmapAn}. In this
case, the optimal choice corresponds to the value of the internal
parameter $\boldsymbol{\beta}$ which minimizes the average KL
divergence to the sources $\boldsymbol{\alpha}$, defined as
\begin{equation}
  \displaystyle
  d(\rho_{\mathrm{src}}|\boldsymbol{\beta}):= \int d \boldsymbol{\alpha} 
  \rho_{\mathrm{src}}(\boldsymbol{\alpha})  D(\boldsymbol{\alpha}|\boldsymbol{\beta})  \ .
\label{eq:annmin}
\end{equation}
Plugging the specific expression for the KL divergence,
i.e. equation~\ref{eq:KLgen}, into this, we obtain an equation for the
stationary points:
\begin{equation}
  \displaystyle
  0 = \parcial{}{\beta_\mu} d(\rho_{\mathrm{src}}|\boldsymbol{\beta})=
  -  
  \ang{\phi_{\mathrm{int}}^\mu}_{\boldsymbol{\beta}} +  
  \int d \boldsymbol{\alpha} \ \rho_{\mathrm{src}}(\boldsymbol{\alpha}) \
  \ang{\phi_{\mathrm{int}}^\mu}_{\boldsymbol{\alpha}}    \ .
\label{eq:annmin2}
\end{equation}
This equation can be interpreted in an alternative way; introducing the ``averaged environment''
\begin{equation}
  \displaystyle
  \bar{P}_{\mathrm{src}}(\mathbf{s}|\rho_{\mathrm{src}}) :=
  \int d \boldsymbol{\alpha} \rho_{\mathrm{src}}(\boldsymbol{\alpha}) 
  P_{\mathrm{src}}(\mathbf{s}|\boldsymbol{\alpha}) \ ,
\label{eq:aveenv}
\end{equation}
the KL divergence respect to
$P_{\mathrm{int}}(\mathbf{s}|\boldsymbol{\beta})$ is
\begin{equation}
  \displaystyle
  D(\bar{P}_{\mathrm{src}}(\cdot|\rho_{\mathrm{src}})|
  P_{\mathrm{int}}(\cdot|\boldsymbol{\beta}) ) 
  = d(\rho_{\mathrm{src}}|\boldsymbol{\beta}) - \int d\boldsymbol{\alpha}
  \rho_{\mathrm{src}}(\boldsymbol{\alpha})
  D(P_{\mathrm{src}}(\cdot|\boldsymbol{\alpha})|\bar{P}_{\mathrm{src}}(\cdot|\rho_{\mathrm{src}}) ) \ .
\label{eq:aveenv2}
\end{equation}
Since the last term on the right hand side does not depend on
$\boldsymbol{\beta}$, the minimization of the KL divergence between the
``averaged environment'' and the internal mapping
$P_{\mathrm{src}}(\mathbf{s}|\boldsymbol{\beta})$ leads to the same result as the
minimization of $d(\rho_{\mathrm{src}}|\boldsymbol{\beta})$ given by
eq.~\ref{eq:annmin}.  In both cases, the Hessian matrix turns out to
be strictly positive (see equation~\ref{eq:statHessian}), and
therefore the eventual extrema are local minima.

In the particular case in which $I=E$, writing
$\phi_{\mathrm{src}}^\mu=\phi_{\mathrm{int}}^\mu=\phi^\mu$, the
optimal internal parameters $\boldsymbol{\beta}$ is given by the
solution of eq.~(\ref {eq:annmin2}), which can be simply written as
\begin{equation}
\displaystyle
\big< \phi^\mu \big>_{\boldsymbol{\beta}} = \int d \boldsymbol{\alpha}
\ 
\rho_{\mathrm{src}}(\boldsymbol{\alpha}) \big< \phi^\mu \big>_{\boldsymbol{\alpha}} \ .
\label{eq:optimal_ann}
\end{equation}

Now, we proceed to study how the distance to the critical point is
modified by an internal representation. That is, we take a narrow
distribution of $\boldsymbol{\alpha}$ values, characterized by a given
average value, $\boldsymbol{\bar{\alpha}}$, at some distance to
criticality, and wonder how does the distance to criticality change
when the optimal internal representation is constructed? Does it grow
or does it generically diminish?

For this purpose, we analyze the local behavior of the internal parameters, taking
a distribution $\rho_{\mathrm{src}}(\boldsymbol{\alpha})$ which is different from zero
only in a small region $U$. We can therefore expand both of the left
and the right hand sides of eq. \ref{eq:optimal_ann} around the mean
value $\bar{\boldsymbol{\alpha}} := \int_U d \boldsymbol{\alpha} \
\rho_{\mathrm{src}}(\boldsymbol{\alpha}) \boldsymbol{\alpha}$, and then we obtain
\begin{equation}
  \displaystyle
  (\beta_\mu-\bar{\alpha}_\mu) =
  \left(\chi^{-1}\right)^{\mu\nu}(\bar{\boldsymbol{\alpha}}) 
  \frac{1}{2} \left. \frac{\partial }{\partial \alpha_\nu}
    \chi^{\gamma\delta}(\boldsymbol{\alpha}) 
  \right|_{\bar{\boldsymbol{\alpha}}} \int_U  d \boldsymbol{\alpha}' 
  \rho_{\mathrm{src}}(\boldsymbol{\alpha}')
  (\alpha'_\gamma - \bar{\alpha}_\gamma)( \alpha'_\delta - \bar{\alpha}_\delta )  \ ,
\label{eq:expansion_ann}
\end{equation}
where $\chi^{\mu\nu}(\boldsymbol{\alpha})$ has been already defined in
eq.~\ref{eq:suscep_mapappa}. This is an equation for the deviation of
the internal parameter $\boldsymbol{\beta}$ respect to
$\bar{\boldsymbol{\alpha}}$. To understand its relation with the
critical point, we rewrite eq. \ref{eq:expansion_ann} as
\begin{equation}
  \displaystyle
  (\boldsymbol{\beta}-\bar{\boldsymbol{\alpha}}) = 
  \chi^{-1}(\bar{\boldsymbol{\alpha}}) \nabla \Omega(\bar{\boldsymbol{\alpha}}),
\label{eq:expansion_ann2}
\end{equation}
where we have defined the scalar field $\Omega(\boldsymbol{\alpha})$ as
\begin{equation}
  \displaystyle
  \Omega(\boldsymbol{\alpha}):= \frac{1}{2}  \int_U  d \boldsymbol{\alpha}' \ \rho_{\mathrm{src}}(\boldsymbol{\alpha}')
  (\boldsymbol{\alpha}' - \boldsymbol{\bar{\alpha}})^T \cdot \chi(\boldsymbol{\alpha}) \cdot ( \boldsymbol{\alpha}' - \boldsymbol{\bar{\alpha}} ) \ .
\label{eq:omega_ann}
\end{equation}

As $\chi(\boldsymbol{\alpha})$ is a positive-definite matrix, from
eq. \ref{eq:omega_ann} we see that $\Omega(\boldsymbol{\alpha})$ is a positive
quantity.  If we introduce a base of eigenvectors of
$\chi(\boldsymbol{\alpha})$, $\mathbf{v}^\gamma(\boldsymbol{\alpha})$ with eigenvalues
$\lambda^\gamma(\boldsymbol{\alpha})$, we get
\begin{equation}
\displaystyle
\Omega(\boldsymbol{\alpha})= \frac{1}{2}\lambda^\gamma(\boldsymbol{\alpha}) \int_U  d \boldsymbol{\alpha}' \ \rho_{\mathrm{src}}(\boldsymbol{\alpha}')
\big[ \mathbf{v}^\gamma(\boldsymbol{\alpha})^T\cdot ( \boldsymbol{\alpha}' - \boldsymbol{\bar{\alpha}} ) \big]^2\ .
\label{eq:omega2_ann}
\end{equation}
At the critical point, at least one $\lambda^\gamma(\boldsymbol{\alpha})$
diverges (in the thermodynamic limit), so $\Omega(\boldsymbol{\alpha})$ has a
maximum at the critical point. Note that
$\mathbf{v}^\gamma(\boldsymbol{\alpha})$ cannot vanish because it is an
unitary vector. Therefore, the gradient of $\Omega$ points to the
critical point, at least if $\bar{\boldsymbol{\alpha}}$ is not too far from it in a such
a way that there are not other local maxima to which the gradient
could be pointing.

Finally, we project both sides of eq. \ref{eq:expansion_ann2} over the
gradient of $\Omega(\bar{\boldsymbol{\alpha}})$:
\begin{equation}
\displaystyle
(\nabla\Omega(\bar{\boldsymbol{\alpha}}))^T\cdot \left(\boldsymbol{\beta}-\bar{\boldsymbol{\alpha}} \right) =
(\nabla\Omega(\bar{\boldsymbol{\alpha}}))^T\cdot\chi^{-1}(\bar{\boldsymbol{\alpha}})\cdot\nabla\Omega(\bar{\boldsymbol{\alpha}}) \ .
\label{eq:omega3_ann}
\end{equation}
As $\chi$ is positive-definite, also its inverse $\chi^{-1}$ is a
positive-definite matrix, and the projection of
$(\boldsymbol{\beta}-\bar{\boldsymbol{\alpha}})$ on the gradient of $\Omega(\bar{\boldsymbol{\alpha}})$
(which points to the critical point) is also positive. 

Consequently, the internal parameters
$\rho_{\mathrm{src}}(\boldsymbol{\alpha})\longrightarrow
\boldsymbol{\beta}$ are closer to the critical point than
$\bar{\boldsymbol{\alpha}}$, indicating that there is an overall drift
towards parameter regions with larger Fisher information, i.e. toward
criticality.

\subsection{Analytical results II: Examples}

\subsubsection{Quenched choice}
We analyze numerically a simple case, inspired by the archetypical
(mean-field) Ising model \cite{Binney-S}, in which
\begin{equation}
  H_{\mathrm{src}}(\mathbf{s}|\boldsymbol{\alpha}) = -\dfrac{N}{2} \alpha_1
  \left(\sum_i\dfrac{s_i}{N} \right)^2 -
  \dfrac{N}{4!} \alpha_2 \left(\sum_i \dfrac{s_i}{N}\right)^4
\label{eq:ising4}
\end{equation}
and
\begin{equation}
 H_{\mathrm{int}}(\mathbf{s}|\boldsymbol{\beta}) = -\dfrac{N}{2}
 \beta_1 
\left(\sum_i \dfrac{s_i}{N}\right)^2 = -\dfrac{1}{N} \beta_1 \sum_{i,j>i}^N s_i s_j + \text{constant}.
\label{eq:ising}
\end{equation}
Numerical factors $(-N/2)$ and $(-N/4!)$ have been introduced for
convenience. The last form of the $H_{\mathrm{int}}$ visually relates this internal representation with the classic problem of Boltzmann learning \cite{Boltzmannmachine-S}.

\begin{figure}[h!]
\centering\includegraphics[width=0.6\textwidth]{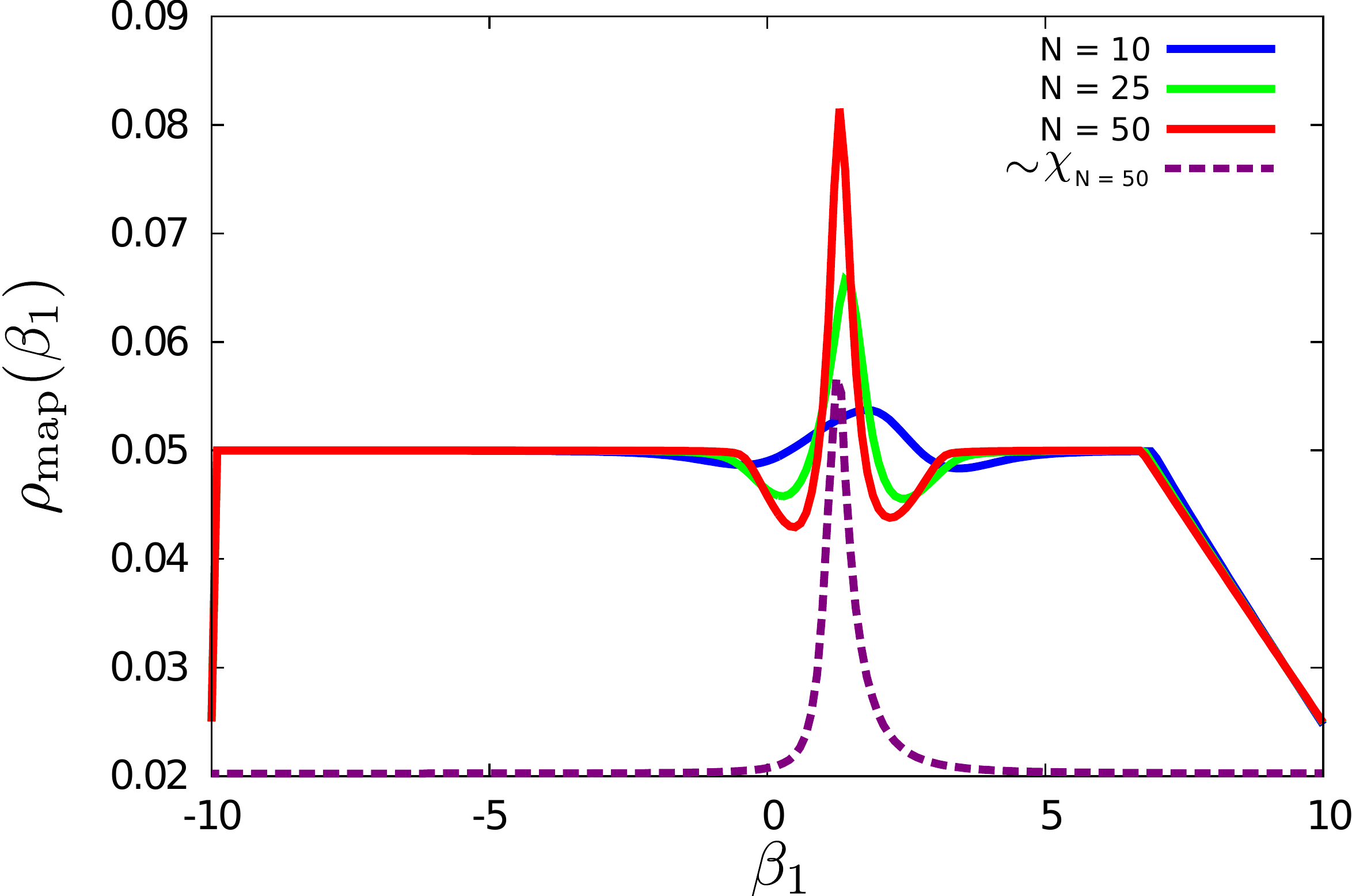}
\caption{{\it Quenched choice:} $\rho_{\mathrm{int}}(\beta_1)$ for
  source and individual parametrizations given in eq.~\ref{eq:ising4}
  and~\ref{eq:ising}, respectively. In this case, we choose
  $\rho_{\mathrm{src}}(\alpha_1,\alpha_2)$ to be uniform in the range
  $[-10,10]\times[-20,20]$. Colored solid lines represent
  $\rho_{\mathrm{int}}(\beta_1)$ for different values of $N$, while
  the dashed line is proportional to the generalized susceptibility of
  the model (as defined by eq.~\ref{eq:suscep_mapappa}, for $N=50$,
  and rescaled for visualization purposes).  The maximum of this
  distribution, which turns out to be located very near the peak of
  the susceptibility, represents the optimal mapping, as given in
  eq.~\ref{eq:optmapQue}. }
\label{fig:quenched}
\end{figure}

As explained in section \ref{sec:quenched}, we take a specific
distribution $\rho_{\mathrm{src}}(\boldsymbol{\alpha})$ --in this case a uniform
distribution in the range $(\alpha_1,\alpha_2)\in[-10,10]\times[-20,20]$-- and
we compute the internal parameters distribution $\rho_{\mathrm{int}}(\boldsymbol{\beta})$. The result
is shown in Fig. \ref{fig:quenched} together with the corresponding
generalized susceptibility, for different values of $N$,

\subsubsection{Annealed model}

\begin{figure}[h!]
\centering\includegraphics[width=0.6\textwidth]{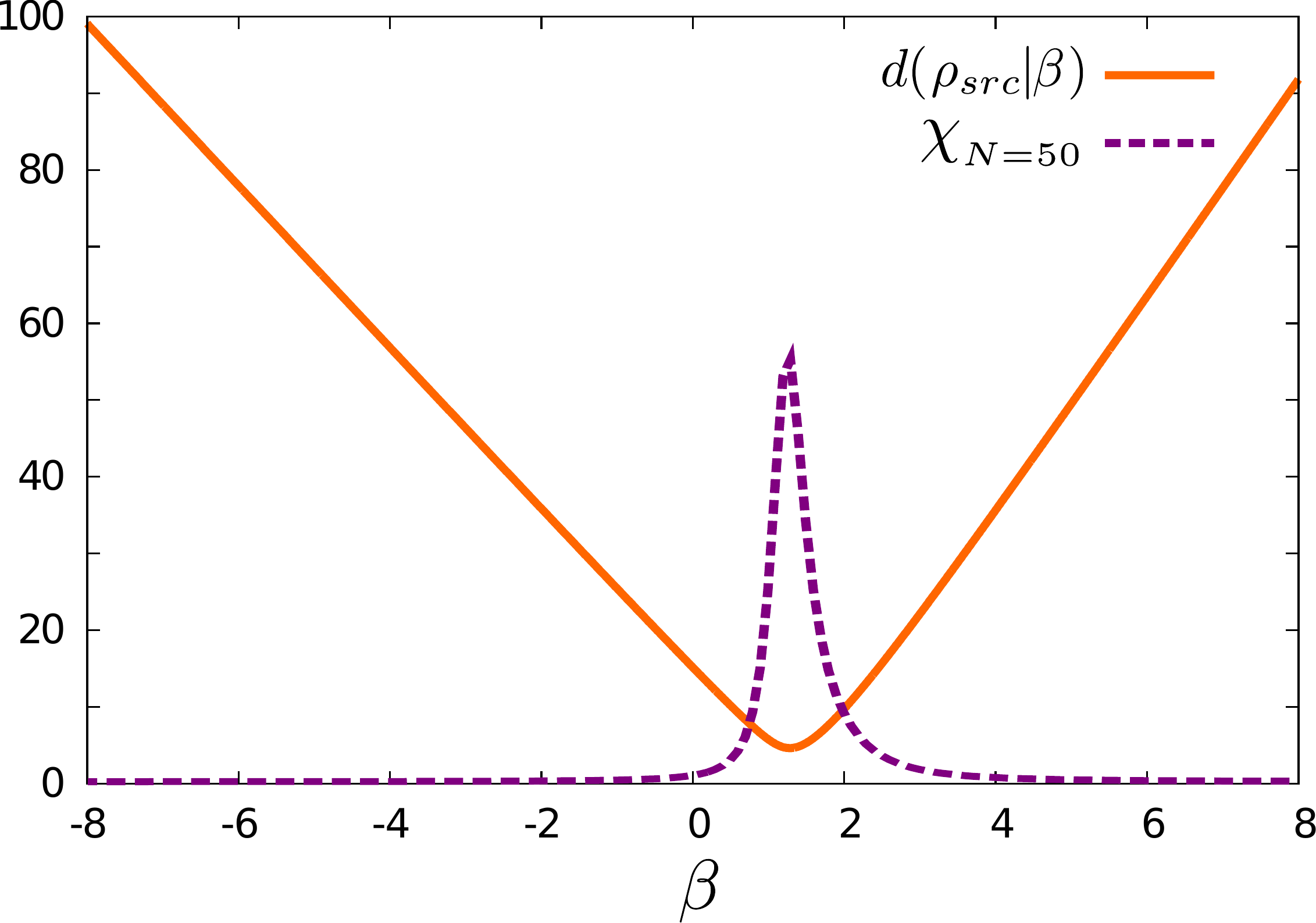}
\caption{{\it Annealed choice.} The orange line is the averaged KL
  divergence $d(\rho_{\mathrm{src}}|\beta)$ of the internal parameters
  $\beta$, defined in eq.~\ref{eq:annmin}. The dashed line is
  proportional to the susceptibility of the internal model (defined in
  eq.~\ref{eq:suscep_mapappa}). The minimum of the distance is located
  close to the critical point (defined by the peak of the generalized
  susceptibility).  In this case both the source and the internal
  representations are modeled by equation~\ref{eq:ising} with $N=50$,
  and the source parameters are uniformly distributed in the range
  $[-10,10]$.  }
\label{fig:annealed}
\end{figure}

We now study a particular case of the theory developed in section
\ref{sec:annealed}. The difference respect to the previous case is
that, here, we compute the mean distance of internal parameters $\boldsymbol{\beta}$
respect to all the sources $\boldsymbol{\alpha}$. Now, both the sources and the
internal representations are modeled by eq.\ref{eq:ising}, each one containing only one
parameter. In this case, we considered $\rho_{\mathrm{src}}(\alpha)$ uniform in the
range $(-10,10)$. The result of the mean distance as a function of
$\beta$ is plotted in Fig.~\ref{fig:annealed}.

\section{Computational models}
\label{sec:evolutionary}

We study the dynamics of two different evolutionary models, both of
them inspired by a genetic algorithm \cite{GA1-S, GA2-S}. The first one consists of an
ensemble of $M$ individuals evolving in a world providing varying
``external'' stimuli. On the other hand, in the Co-evolutionary Model,
individuals evolve with a similar algorithm but instead of having to
cope with an external environment the community itself plays the role
of the external world, i.e. individuals need to be able to respond to the state of other
similar individuals (which play also the role of sources).  We have studied different variants of this
general model (see below).

\subsection{Co-evolutionary Model}

Here we discuss the type of evolutionary model, in which every
individual receives stimuli from its surrounding world, which is
nothing but the set of the other individuals in the community.  More
specifically: the $k$-th agent of the community is described by a
probability distribution $P_{\mathrm{int}}(\mathbf{s}|\boldsymbol{\beta}^k)\propto
\exp\{-H_{\mathrm{int}}(\mathbf{s}|\boldsymbol{\beta}^k)\}$, depending on parameters
$\boldsymbol{\beta}^k$.

Starting with an ensemble of $M$ individuals whose coupling parameters
are extracted from an arbitrary distribution, $p(\boldsymbol{\beta},t=0)$, the
evolutionary dynamics proceeds as follows:

\begin{enumerate}

\item At each time step, two individuals, $i$ and $j$, are randomly selected.

\item Their relative fitnesses $f_i^{(j)}$ and $f_j^{(i)}$, defined as the complementary
  KL divergences from the one to the other:
\begin{equation}
f_i^{(j)}=D(\boldsymbol{\beta}^i|\boldsymbol{\beta}^j) = \sum_{\mathbf{s}}
P_{\mathrm{int}}(\mathbf{s}|\boldsymbol{\beta}^i)\log\dfrac{P_{\mathrm{int}}(\mathbf{s}|\boldsymbol{\beta}^i) }{P_{\mathrm{int}}(\mathbf{s}|\boldsymbol{\beta}^j)}, \qquad 
f_j^{(i)}=D(\boldsymbol{\beta}^j|\boldsymbol{\beta}^i) =
\sum_{\mathbf{s}}  P_{\mathrm{int}}(\mathbf{s}|\boldsymbol{\beta}^j)\log\dfrac{P_{\mathrm{int}}(\mathbf{s}|\boldsymbol{\beta}^j)
}{P_{\mathrm{int}}(\mathbf{s}|\boldsymbol{\beta}^i)},
\end{equation}
where --as the KL divergence is not symmetric-- $f_i^{(j)} \neq f_j^{(i)}$
unless $\boldsymbol{\beta}^i=\boldsymbol{\beta}^j$.

\item One of the two individuals --selected with probability
  proportional to its relative fitness-- creates an offspring, while the other
  one is removed from the community.

\item Offspring mutate with a probability $\nu$, modifying its
  parameters from $\boldsymbol{\beta}$ to
  $\boldsymbol{\beta}\rightarrow \boldsymbol{\beta}+
  \boldsymbol{\xi}$, where $\boldsymbol{\xi}$ is randomly chosen from
  a multivariate Gaussian distribution with zero mean and deviation
  $\boldsymbol{\sigma}$.

\item Time is updated to $t\rightarrow t + 1/M$.

\item Another couple of individuals $i'$ and $j'$ is picked, and the
  process is iterated.
\end{enumerate}

To compute the stationary distribution of parameters
$p(\boldsymbol{\beta})\equiv p(\boldsymbol{\beta},t\rightarrow\infty)$, we iterate
$T_i$ time steps and then perform measurements during $T_f-T_i$ steps.
Results are averaged over $R$ realizations of the evolutionary
process.

We now present some specific realizations of these general co-evolutionary
rules. The ``internal'' probability distributions of a single
individual are taken to as
\begin{equation}\label{magnetic}
  H_{\mathrm{int}}(\mathbf{s}|\boldsymbol{\beta}) = - \dfrac{N}{2} \beta_1
  \left(\sum_i \dfrac{s_i}{N} \right)^2 - N \beta_2 \left(\sum_i
    \dfrac{s_i}{N} \right)
\end{equation}
with a linear and a quadratic coupling.  We will also discuss the
simpler, particular case in which $\beta_2$ vanishes (i.e. only the
quadratic coupling is present), as well as the quadratic-quartic
case:
\begin{equation}\label{4spins}
  H_{\mathrm{int}}(\mathbf{s}|\boldsymbol{\beta}) = -\dfrac{N}{2} \beta_1
  \left(\sum_i\dfrac{s_i}{N} \right)^2 - \dfrac{N}{4!} \beta_2 \left(\sum_i \dfrac{s_i}{N}\right)^4
\end{equation}

Figure 2 of the main text, shows the distribution of individual
parameters, $p(\boldsymbol{\beta})$ for the linear-quadratic case,
eq.~\ref{magnetic}, for three different initial distributions and
different transient periods, together with the final stationary
distribution which is the same for all of them. Numerical parameters
are summarized in Table \ref{tab:parameters-e2}.
\begin{table}[h]
\begin{center}
\begin{tabular}{l|r}
  {\bf Parameter} & {\bf Value}\\\hline
  $N$ & 100\\
  $M$ & 100\\
  $\nu$ & 0.1\\
  $\sigma$ & 0.1\\
  $\sigma_{1},\sigma_{2}$ & $1\sigma, 0.1\sigma$\\
    $T_i$ & $10^4$\\
  $T_f-T_i$ & $10^5$\\
  Init. Distribution & $N(-3,0.25)\cdot N(-0.25,0.05)$\\
& $N(3,0.25)\cdot N(0.25,0.05)$\\
& $U([-4,4]\times[-0.8,0.8])$\\
$R$ & 100 (10000 for transients) \\
 \end{tabular}
 \caption{Parameter values in the simulation of the Co-evolutionary
   Model in Fig. 2 of the main text. See also Table
   \ref{tab:parameters-e1}.}
\label{tab:parameters-e2}
\end{center}
\end{table}

\subsubsection{Results}

\begin{figure}[h!]
\centering\includegraphics[width=0.7\textwidth]{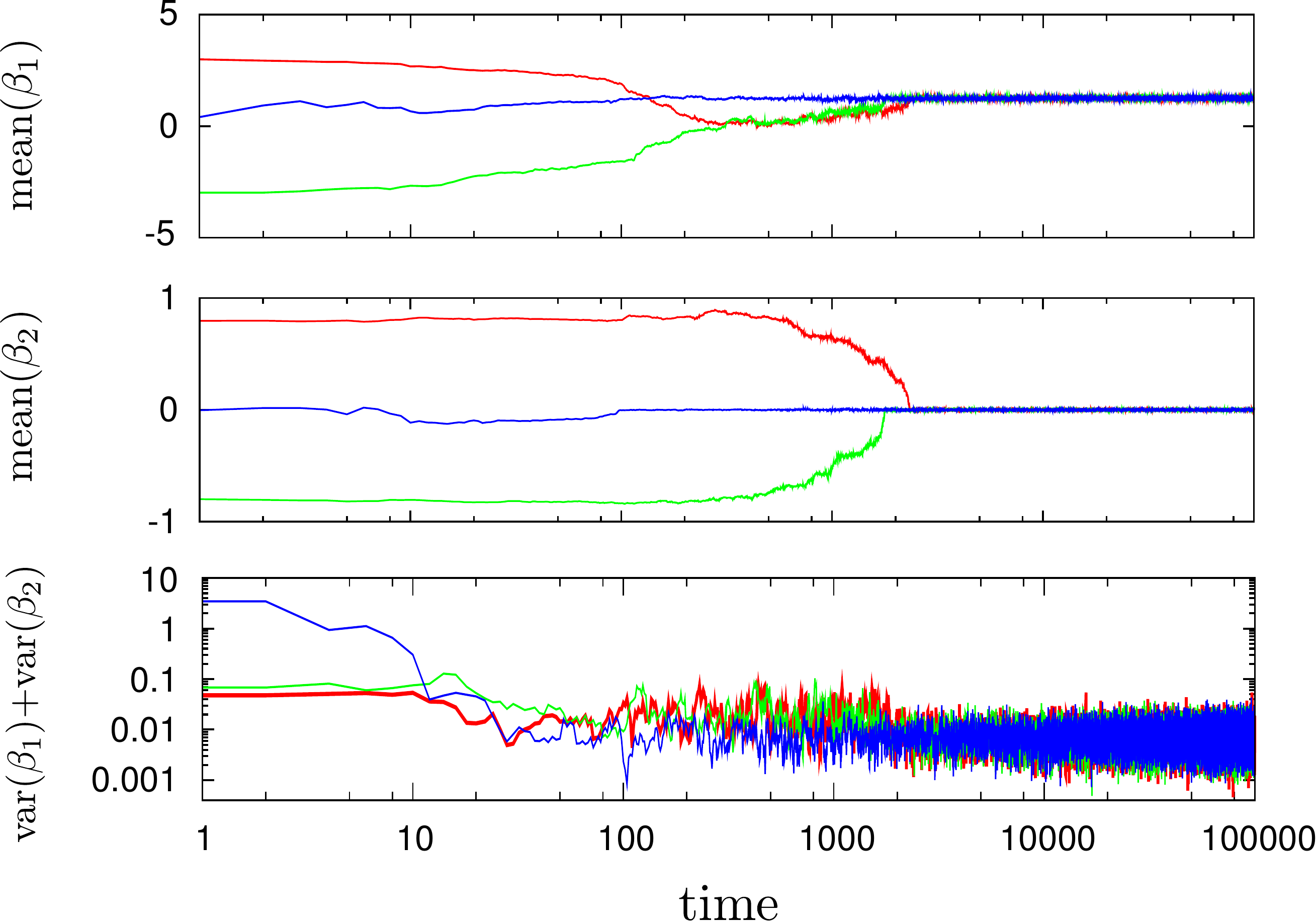}
\caption{{\it Time evolution in the Co-evolutionary Model} (in the
  linear-quadratic case, i.e.  eq.~\ref{magnetic}). Plot of the mean
  values $\ang{\beta_{1,2}}:=\frac{1}{M}\sum_{k=1}^M \beta_{1,2}^k$
  (top two panels) of the community, and parameter variance (bottom
  panel) for three different realizations started with different
  initial conditions $p(\boldsymbol{\beta},t=0)$: Gaussian
  distributions with two different averages and variances,
  $N(-3,0.25)N(-0.8,0.05)$ (red line) and $N(3,0.25)N(0.8,0.05)$
  (green line), and uniform distribution in the range
  $[-4,4]\times[-0.8,0.8]$ (blue line).}
\label{fig:e2_stat}
\end{figure}

\begin{figure}[h!]
\centering\includegraphics[width=0.8\textwidth]{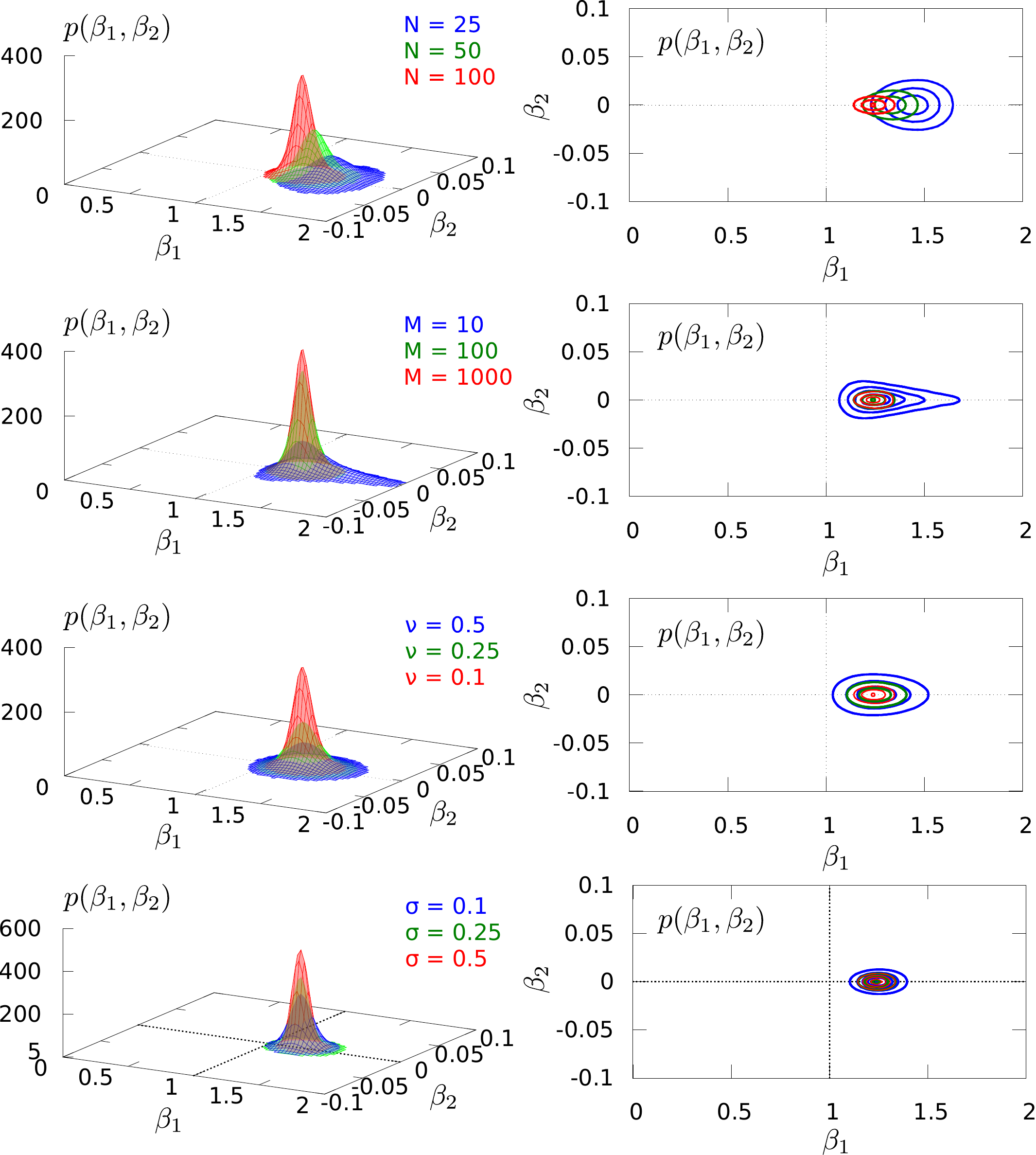}
\caption{{\it Parameter dependence in the Co-evolutionary Model:}
  stationary distribution $p(\beta_1,\beta_2)$ as a function of
  parameters (for the linear-quadratic case,
  eq.~\ref{magnetic}). Different colored lines in each plot correspond
  to different values of $N$, community sizes $M$, mutation parameters
  $\nu$, and $\sigma$. For larger communities sizes the stationary
  distribution becomes sharper. Parameter values are listed in
  Table~\ref{tab:parameters-e2} (unless otherwise specified).}
\label{fig:e2_parameters}
\end{figure}

\begin{figure}[h!]
\centering\includegraphics[width=0.6\textwidth]{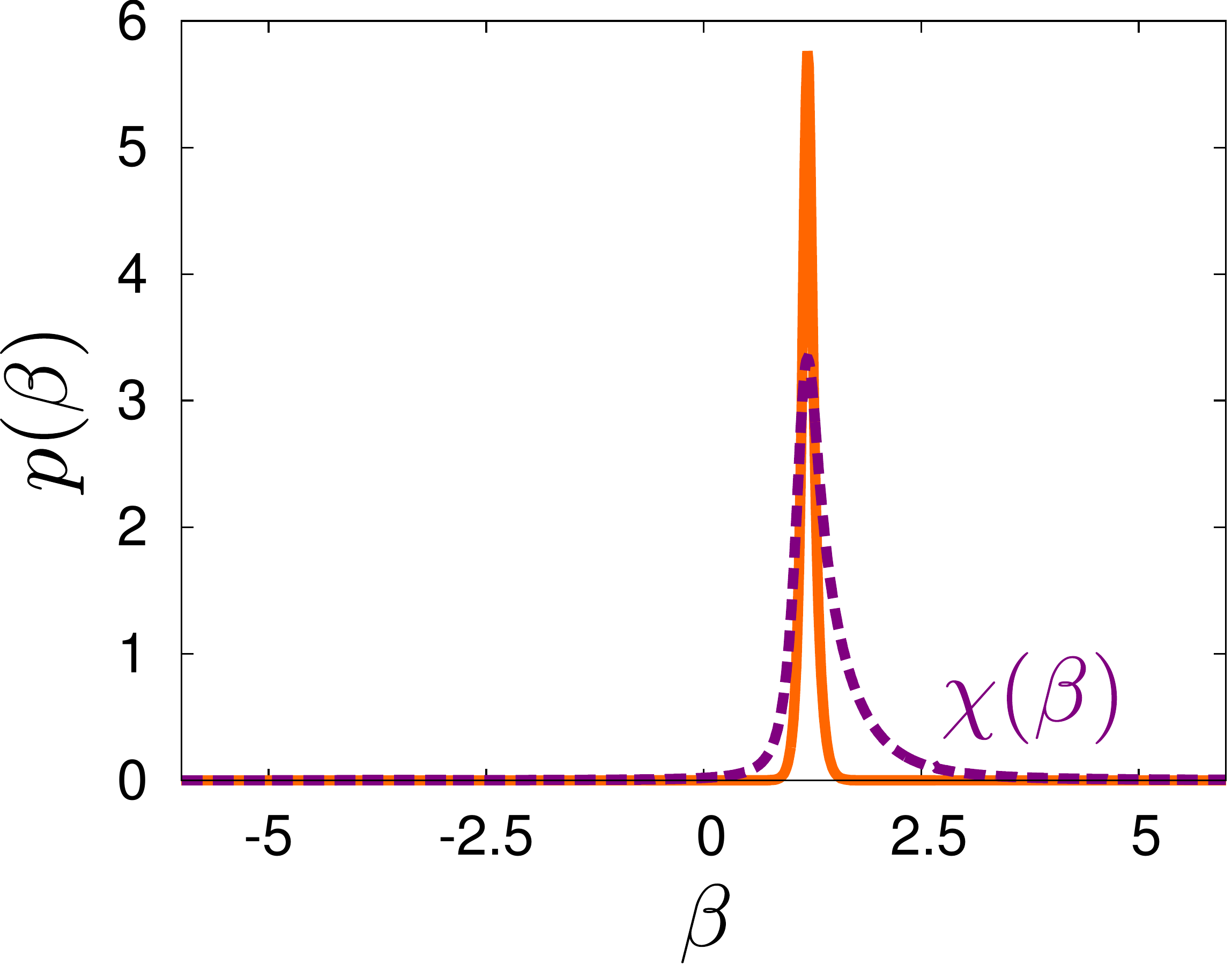}
\caption{{\it Stationary distribution $p(\beta)$ in the
    Co-evolutionary Model, eq. \ref{eq:ising}:} we compare the
  stationary distribution (orange line) with the generalized
  susceptibility (purple dashed line).  As in the example in the main
  text, the individual parameters converge to the neighborhood of the
  peak of the Fisher information.   Parameters are set to
  $N=100$, $M=100$, $\nu=0.1$, $\sigma=0.1$.  }
\label{fig:e2_J}
\end{figure}

\begin{figure}[h!]
\centering\includegraphics[width=0.9\textwidth]{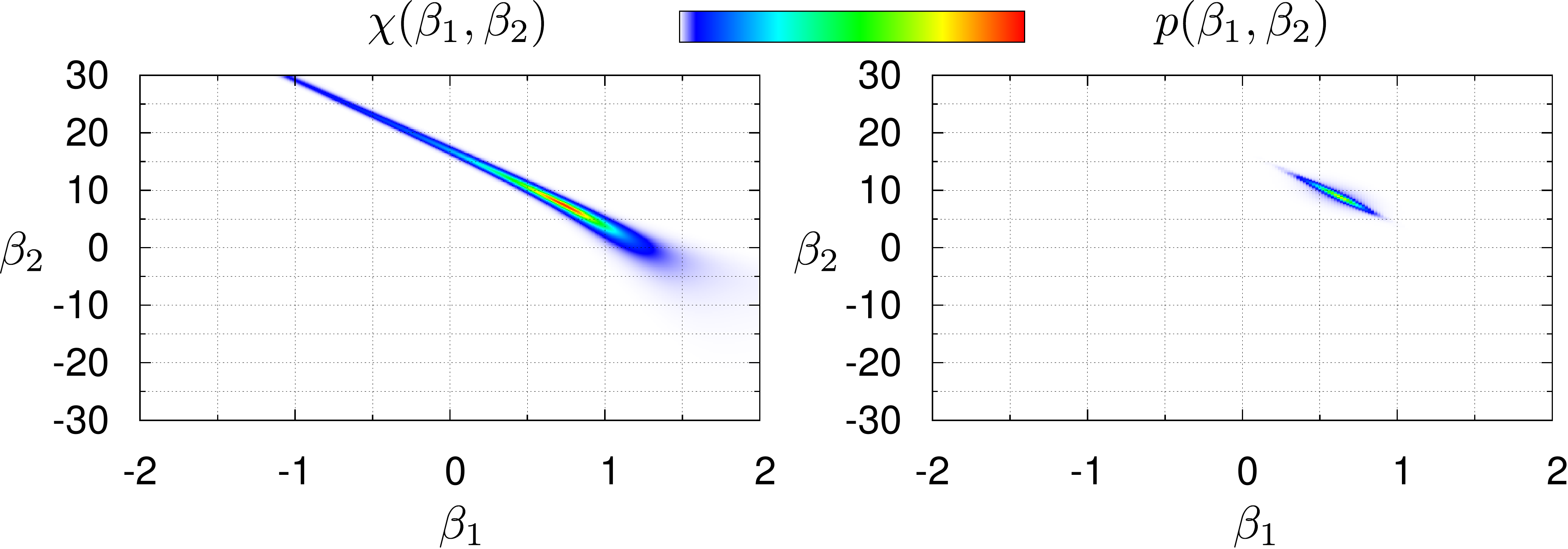}
\caption{{\it Stationary distribution $p(\beta_1,\beta_2)$ in the
    Co-evolutionary Model, eq. \ref{4spins}:} as above, we compare the
  stationary distribution (right panel) with the generalized
  susceptibility (left panel).  Again, the community evolves toward
  the global maximum of the Fisher information. Parameter values:
  $N=100$, $M=100$, $\nu=0.1$, and
  $\sigma_{1}=\sigma_{2}=0.1$.  }
\label{fig:e2_JL}
\end{figure}

As for the Evolutionary Model, we first study the stationarity of the
final distribution of parameters $\boldsymbol{\beta}$ in the community
and its dependence on number of spins, $N$, number of individuals,
$M$, the mutation probability, $\nu$, and the deviation of the
mutations, $\sigma$. Results for the case of a community with
individuals whose probability distribution is characterized by the
linear-quadratic case, eq. \ref{magnetic} are summarized in
Figs. \ref{fig:e2_stat} and \ref{fig:e2_parameters}, whereas
Figs. \ref{fig:e2_J} and \ref{fig:e2_JL} correspond to the
one-parameter case (eq. \ref{eq:ising}) and the quadratic-quartic case
(eq. \ref{4spins}), respectively.

\newpage
\subsection{Co-evolutionary Model with K-body interactions}

Are the previous results affected if a larger number of individuals is
allowed to interact at each time step?

To answer this question, here we study a variant of the Co-evolutionary Model in which
$K$ individuals $(i_1,...,i_K)$ are randomly picked at each time step,
and they compete among themselves; the probability to die is
proportional to its (normalized) average KL divergence to the
remaining ones, i.e.
\[
P_{\mathrm{kill}}(i_k) = \dfrac{ \displaystyle \sum_{l=1}^{K}
  D(\boldsymbol{\beta}^{i_{l}} | \boldsymbol{\beta}^{i_k} ) }{ \displaystyle
  \sum_{m=1}^K \sum_{l=1}^{K} D(\boldsymbol{\beta}^{i_{l}} | \boldsymbol{\beta}^{i_m}
  ) },
\]
and then it is replaced by a copy of one of the remaining $K-1$
individuals (and mutations are introduced with probability $\nu$).

\begin{figure}[h!]
  \centering\includegraphics[width=0.9\textwidth]{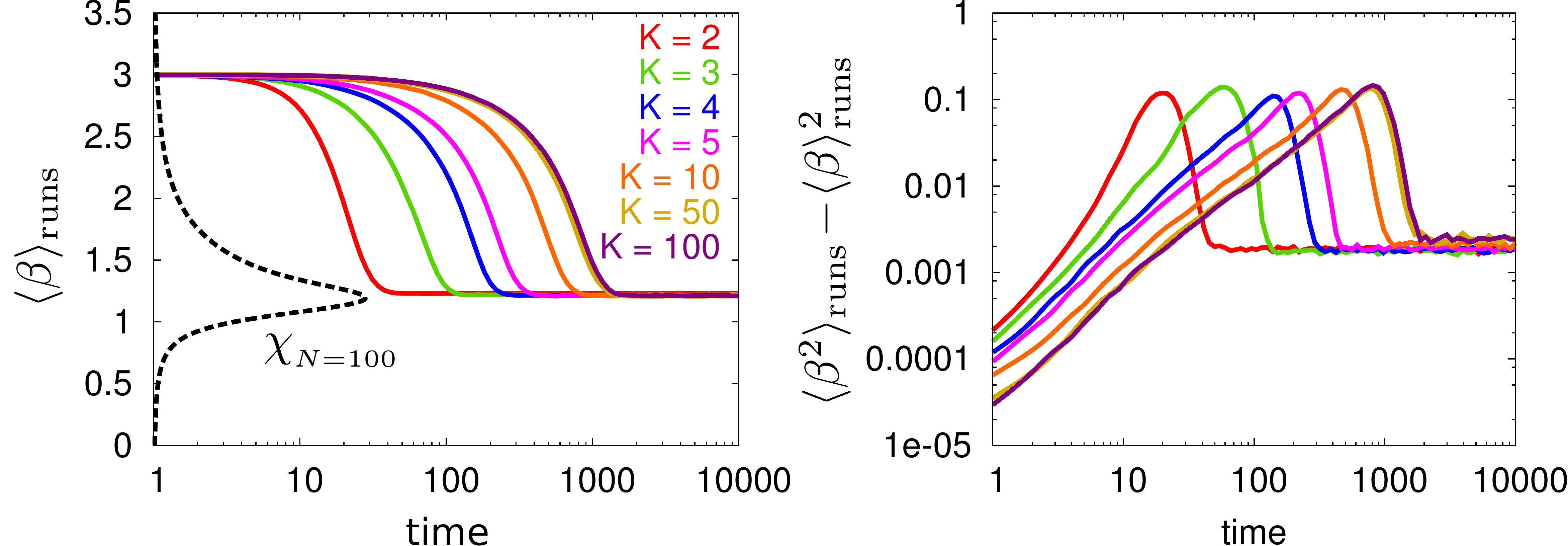}
  \caption{{\it Time evolution in the Co-evolutionary Model with
      K-body interactions} (with parametrization of
    eq.~\ref{eq:ising}). Solid lines represent the time evolution of
    mean values $\ang{\beta}:=\int d\beta \ p(\beta,t) \beta$.  The
    relaxation to the stationary state depends on the effective number
    of individuals $K$ with which each single agent interacts. The
    larger the value of $K$ the larger the relaxation time. The
    community evolves very close to the maximum of the generalized
    susceptibility (plotted with dashed lines).  The initial condition
    is $\beta=3$ for all of the individuals and parameters are
    $N=100$, $M=100$, $\nu=1$ and $\sigma=0.1$.  }
\label{fig:e2_many_bodies}
\end{figure}

We have implemented the simulation with parametrization of eq.~\ref{eq:ising}. Results are summarized in Fig.  \ref{fig:e2_many_bodies} where we plot
the mean value of the parameter over the entire community and $10^3$
realizations of the same initial condition. It can be seen that the
time to reach stationarity increases with $K$. When $K$ increases, the
drift which moves the system towards the criticality is lower. This is
related to the fact that, by averaging over more and more individuals,
the source effectively becomes more and more homogeneous.

\subsection{Co-evolutionary Model with complex internal networked topologies}
\label{sec:E2-III}

We now scrutinize a different variant of the model in which the internal probability distribution of each
individual/agent is not a ``mean-field'' one, in the sense that every
$s_i$ variable is coupled to all others, but instead, possible
interactions are encoded in a network, such that each $s_i$ interacts
only with other $s_j$ directly connected to it, i.e. for which the
{\it adjacency matrix}, element $a_{ij} \neq 0$.

The evolutionary dynamics is as above, with the only difference that
now the structure of the probability characterizing each individual is
as follows:
\begin{itemize}

\item Given $N$ spin variables, we generate a fixed adjacency matrix of
  interactions $\widehat{a}$. The
  probability to find a certain configuration $\mathbf{s}$ in the $k$-th individual is
  $P_{\mathrm{int}}^{\widehat{a}}(\mathbf{s}|\beta^k)\propto
  \exp\{-H_{\mathrm{int}}^{\widehat{a}}(\mathbf{s}|\beta^k)\}$ with
\begin{equation}
H_{\mathrm{int}}^{\widehat{a}}(\mathbf{s}|\beta^k) =
-\beta^k\dfrac{1}{N}\sum_{i,j>i}^{N} a_{ij} s_i s_j.
\label{eq:e2-d}
\end{equation}

\item The system is iterated as in the Co-evolutionary Model, leaving $\widehat{a}$
  fixed in time and identical for all individuals.
\end{itemize}
As the structure of $\widehat{a}$ is an arbitrary one, the calculation
of the distances between distributions needs to be explicitly computed
by summing over the $2^N$ possible states (which severely limits the
maximum size in computer simulations; $N\sim20$).

\begin{figure}[!h]
\centering \includegraphics[width=\textwidth]{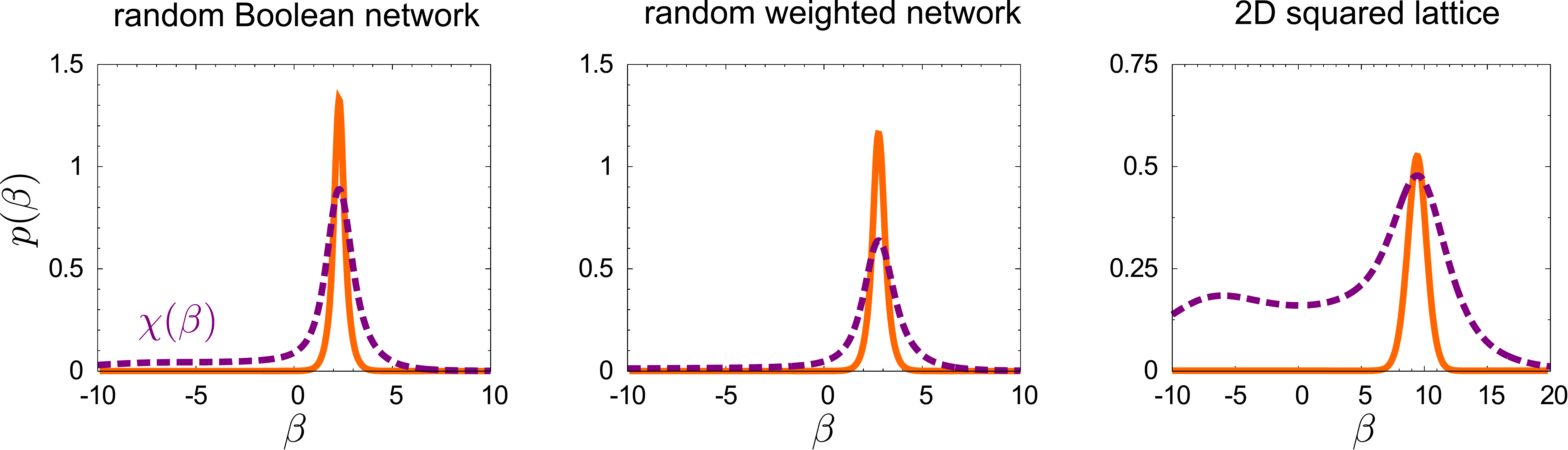}
\caption{Stationary parameter distributions for three different
  network structures $\widehat{a}$ of $N$ nodes. Three cases are
  studied: \textbf{(left) Random Boolean network}: connections are not
  weighted, i.e.  $a_{ij}=a_{ji}=\{0,1\}$, with mean connectivity
  $N/2$, \textbf{(center) Random weighted network}: in this case,
  $a_{ij}=a_{ji}=\eta$, where $\eta$ is a random number between $0$ and
  $1$\textbf{(right) Regular 2D lattice} with periodic bounding
  conditions. In all cases, parameters have been set to $\nu=0.1$,
  $\sigma=0.5$, $M=100$. In the first two examples, $N=20$, while
  $N=25$ for last one. In dashed line, the generalized
  susceptibility has been plotted and re-scaled for visual comparison.
}
\label{fig:nets}
\end{figure}

Results for the stationary distribution of parameters $p(\beta)$ and
different types of network architectures, together
with the corresponding curves of generalized susceptibilities
computed as
\begin{equation}
  \chi^{\widehat{a}}(\beta) =
  \left\langle\left(\dfrac{1}{N}\sum_{i,j>i}^{N} a_{ij} s_i
      s_j\right)^2\right
  \rangle_{P^{\widehat{a}}(\mathbf{s}|\beta)} - \quad
  \left\langle\dfrac{1}{N}\sum_{i,j>i}^{N} a_{ij} s_i s_j\right\rangle_{P^{\widehat{a}}(\mathbf{s}|\beta)}^2
\end{equation}
are shown in Fig. \ref{fig:nets}.  In all cases, the main result of
this paper holds: the resulting internal parameters distributions peak around the
critical point.

\subsection{Evolutionary Model}

We model a community of individuals receiving external stimuli from
an outer and heterogeneous environment. We describe every source of
the environment by a generic  distribution, as parametrized by equation~\ref{eq:Hsrc}
with
\begin{equation}
 H_{\mathrm{src}}(\mathbf{s}|\boldsymbol{\alpha}) =  \sum_{\mu}^{E} \alpha_\mu \phi^\mu_{\mathrm{src}}(\mathbf{s}),
\end{equation}
where the parameters $\boldsymbol{\alpha}$ are drawn from the distribution
$\rho_{\mathrm{src}}(\boldsymbol{\alpha})$.

In the community, each agent has a representation of the observed
source (constructed as explained in previous sections). The internal state of the
$k$-th agent is modeled with an internal system described by
equation~\ref{eq:Hmap}, with
\begin{equation}\label{eq:Hint}
 H_{\mathrm{int}}(\mathbf{s}|\boldsymbol{\beta}^k) =  \sum_{\mu}^{I} \beta_\mu^k \phi^\mu_{\mathrm{int}}(\mathbf{s}),
\end{equation}

We consider the
following dynamics:

\begin{enumerate}
\item We start with $M$ individuals each one equipped with some
  initial parameter set extracted from some arbitrary (broad)
  distribution.  $p(\boldsymbol{\beta},t=0)$.

\item At every time step, we generate $S$ external sources,
  $\{\boldsymbol{\alpha}^u\}_{u=1,...,{S}}$, from the distribution
  $\rho_{\mathrm{src}}(\boldsymbol{\alpha})$.

  \item We compute the average KL divergence of every individual's internal representation to the external sources
\begin{equation}
  d(\{\boldsymbol{\alpha}^u\}|\boldsymbol{\beta}^k) := \dfrac{1}{{S}} \sum_{u=1}^{S}
\sum_{\mathbf{s}}  P_{\mathrm{src}}(\mathbf{s}|\boldsymbol{\alpha}^u)\log
\dfrac{P_{\mathrm{src}}(\mathbf{s}|\boldsymbol{\alpha}^u)}{P_{\mathrm{int}}(\mathbf{s}|\boldsymbol{\beta}^k)}.
\label{eq:e1_distance}
\end{equation}

\item One of the individuals of the community is removed with a
  probability proportional to its average KL divergence
\begin{equation}
P_{\mathrm{kill}}(k) = \dfrac{d(\{\boldsymbol{\alpha}^u\}|\boldsymbol{\beta}^k)}{\sum_{l}d(\{\boldsymbol{\alpha}^u\}|\boldsymbol{\beta}^{l}) }
\end{equation}
and it is replaced by a copy of another individual (offspring), which is picked
randomly with uniform probability.

\item The offspring inherits its parameter set from its parent or,
  instead, mutates with a probability $\nu$, altering the original
  parameter set, $\boldsymbol{\beta}\rightarrow
  \boldsymbol{\beta}+\boldsymbol{\xi}$, where $\boldsymbol{\xi}$ is a random
  Gaussian number of zero mean and deviation $\boldsymbol{\sigma}$.

\item Time is incremented as $t\rightarrow t + 1/M$.

\item Another set of parameters $\{\boldsymbol{\alpha}^u\}_{u=1,...,S}$ is
  generated from $\rho_{\mathrm{src}}(\boldsymbol{\alpha})$, and the process is
  iterated.
\end{enumerate}

We are interested in measuring the stationary distribution of the
individual parameters, $p(\boldsymbol{\beta})\equiv p(\boldsymbol{\beta},t\rightarrow\infty)$ when
$t\rightarrow\infty$ (we start measuring at some time $T_i$ and stop
at time $T_f$), for which the distribution is averaged over $R$
independent realizations of the initial distribution $p(\boldsymbol{\beta},
t=0)$.

We have simulated the simple case in which both the external sources
and internal representations correspond to the simple choice given in
equation~\ref{eq:ising4} with $\alpha_2=0$ and equation~\ref{eq:ising}
respectively.
Similar results can be obtained for other parametrization of sources
and internal representations, for instance by considering equation~\ref{eq:ising4} with
non-vanishing
$\alpha_2$.

\subsubsection*{Numerical computation of the Kullback-Leibler divergence}
We explore the dependence of the stationary distribution $p(\beta)$ on
parameter values (see Table \ref{tab:parameters-e1}).

\begin{table}[h]
\begin{center}
\begin{tabular}{l|r}
  {\bf Parameter} & {\bf Value}\\\hline
  $N$ & 100\\
  $M$ & 100\\
  $S$ & 10\\
  $\nu$ & 0.1\\
  $\sigma$ & 0.1\\
  $\rho_{\mathrm{src}}(\alpha_1)$ & $U([-10,10])$\\
\begin{footnotesize}(not used in Fig. 3 of main text)\end{footnotesize} & \\
  $T_i$ & $10^4$\\
  $T_f-T_i$ & $10^5$\\
$R$ & 100\\
 \end{tabular}
 \caption{\textit{Parameters of the simulation of the Evolutionary Model in Fig. 3
     of the main text and Fig. \ref{fig:e1}:} $N$ is the number of
   spins composing each of the individuals, $M$ is the community size,
   $S$ is the number of stimuli received in every interaction with the
   environment, $\nu$ is the mutation probability, $\sigma$ is the
   deviation of the mutated offspring, $T_i$ and $T_f$ are the
   initial and final time steps used for the measure and $R$ is the
   number of independent realizations.}
\label{tab:parameters-e1}
\end{center}
\end{table}

For sufficiently large values of $N$, the sum in
eq. \ref{eq:e1_distance} cannot be explicitly compute because of the
diverging large number of states, $2^N$. However, since
equations~\ref{eq:ising} and~\ref{eq:ising4} depend on $\mathbf{s}$ only
through the magnetization $m = \sum_i s_i /N$,
\begin{equation}
  H_{\mathrm{src}}(\mathbf{s}|\boldsymbol{\alpha}) = -\dfrac{N}{2} \alpha_1
  m^2
\label{eq:isingmf4}
\end{equation}
and
\begin{equation}
 H_{\mathrm{int}}(\mathbf{s}|\boldsymbol{\beta}) = -\dfrac{N}{2} \beta_1 m^2,
\label{eq:isingmf}
\end{equation}
we can compute the sum as follows. Defining $\Gamma(m)$ as the number
of states with $\sum_i s_i /N=m$, and using the Stirling
approximation, one readily obtains:
\begin{equation}
  \Gamma(m) = \binom{N}{\frac{N(1+m)}{2} } \underset{N\gg1}{=}
  \exp\left\{-N\left(\dfrac{1+m}{2}\log{\dfrac{1+m}{2}} + \dfrac{1-m}{2}\log{\dfrac{1-m}{2}}\right)\right\}.
\end{equation}
Then, the KL divergence can be computed as
\begin{equation}
  D(\boldsymbol{\alpha}|\boldsymbol{\beta}) \underset{N\gg1}{=} \int_{-1}^1 \Gamma(m)
  \hat{P}_{\mathrm{src}}(m|\boldsymbol{\alpha})\log\dfrac{\hat{P}_{\mathrm{src}}(m|\boldsymbol{\alpha})}{\hat{P}_{\mathrm{int}}(m|\boldsymbol{\beta})} dm.
\end{equation}
The same type of approximation can be used to calculate
the normalization of $P_{\mathrm{src}}$ and $P_{\mathrm{int}}$.


\subsubsection{Results}

\begin{figure}[h!]
\centering\includegraphics[width=0.8\textwidth]{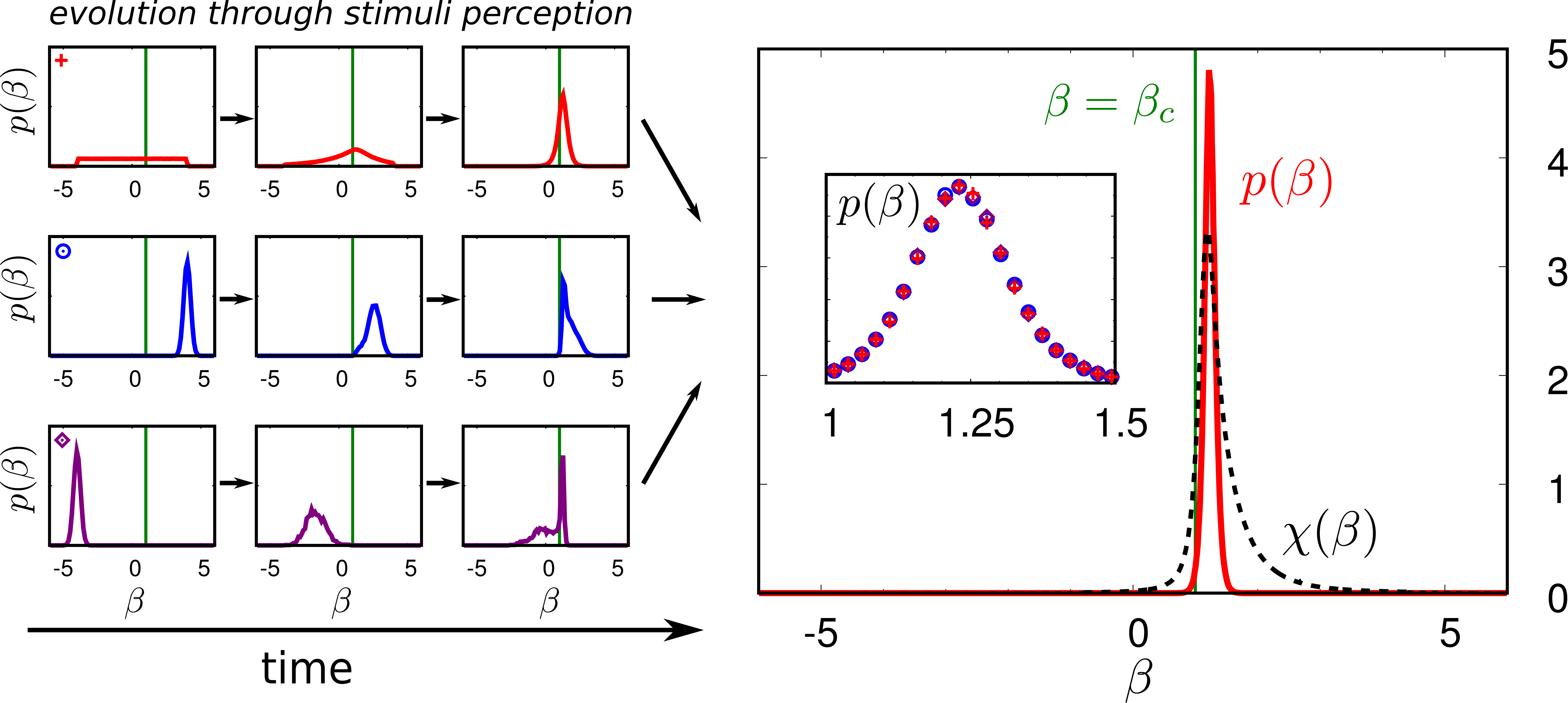}
\caption{{\it Time evolution in the Evolutionary Model }(with parametrization of
  eq. \ref{eq:ising4} with $\alpha_2=0$ and eq. \ref{eq:ising}): Panels on
  the left represent the evolution of three different initial
  distributions of the parameter $\beta$ in the community. In all cases
  the environment is described by the uniform distribution of
  parameters $\rho_{\mathrm{src}}(\alpha)=U([-10,10])$. The distributions converge
  to the same stationary state --points and red line on the right
  panel--, which is peaked at the maximum of the generalized
  susceptibility (dashed line curve).  The red line corresponds to an
  initial uniform distribution $U([-4,4])$, and blue and purple lines
  to Gaussian distributions $N(4,0.25)$ and $N(-4,0.25)$,
  respectively. Parameters are the same as in Table
  $\ref{tab:parameters-e1}$, and $R=10^4$ independent realizations.}
\label{fig:e1}
\end{figure}

Having computed the KL as a function of parameter values we can
iterate the evolutionary dynamics, as described before.  Figure
\ref{fig:e1} illustrates that starting from different initial
conditions $p(\boldsymbol{\beta}, 0)$ after some (sufficiently long) times
the ensemble of individuals converges to a unique steady state
$p(\boldsymbol{\beta}, t \rightarrow \infty)$. The resulting distribution is
sharply peaked very near the critical point, at the very same
location at which the Fisher information or generalized susceptibility
peaks.  This peak approaches the critical point $\beta=\beta_c$ in the limit
$N\rightarrow\infty$.

\begin{figure}[h!]
\centering\includegraphics[width=0.8\textwidth]{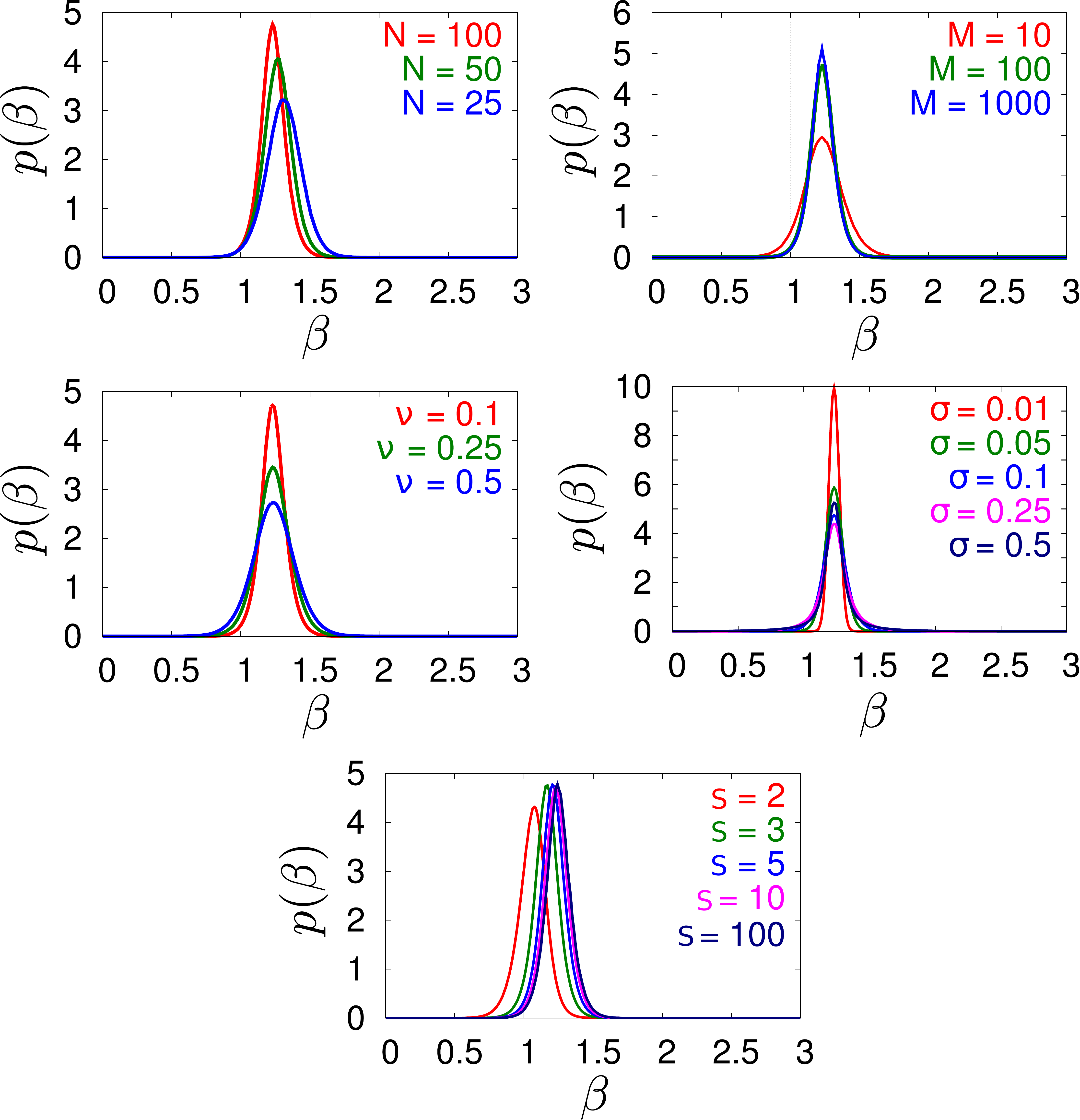}
\caption{{\it Dependence on parameters in the Evolutionary Model:} Stationary
  distribution $p(\beta)$ as a function of diverse parameters; different
  colors in each plot stand for different values of (from top to
  bottom and from left to right): $N$, community size $M$, mutation
  probability $\nu$, mutation deviation $\sigma$, and number of
  external sources $S$. Unless otherwise stated, other parameters
  take the same values as in Table~\ref{tab:parameters-e1}. The dashed
  lines indicate the critical point location (in the limit
  $N\to\infty$) }
\label{fig:e1_parameters}
\end{figure}

First, we proceed to analyze the relaxation of the initial
distribution of parameters $p(\boldsymbol{\beta}, t=0)$ at the
stationary one $p(\boldsymbol{\beta})$, and its dependence on the initial
condition. These results are plotted in Fig.~\ref{fig:e1}. Fig.~\ref{fig:e1_parameters} illustrates the dependence of the
results on parameters.

The main conclusions are:
\begin{itemize}
\item  The system becomes
closer and closer to the critical point as the system-size $N$ is enlarged.
\item
The distribution reaches an asymptotic shape as the ensemble size
grows.
\item The distribution becomes sharper for smaller mutation rates.
\item The distribution becomes much sharper for small mutation
  variances.
\item The distribution reaches an asymptotic shape as the number of
  external sources is increased.
\end{itemize}

\section{Effective criticality and heterogeneity of the environment}

As we have shown in section \ref{sec:annealed}, minimizing the KL
divergence to the ``averaged environment'' --i.e. the distribution of
sources resulting by averaging over different environmental
parameters-- leads to the same result as minimizing the mean KL
divergence to the sources, which is what we implement in the
simulations. Therefore, agents seeing a complex ``averaged
environment'' tend to become critical.

It has been recently suggested that marginalizing a distribution over
some parameters can generally lead to effective critical distributions
~\cite{Schwab2013-S}. Thus it may not be surprising that individuals
tune their parameters near the criticality to minimize the KL
divergence with respect to such a critical environment.

Here we show that our results cannot be generically explained in terms
of this phenomenon. For the cases in which the individuals end up near
criticality, the averaged environment is not necessarily critical.  To
illustrate this, we consider each of the pools of sources used in
Fig. 3 of the main text to illustrate the evolution of agents in the
presence of complex environments.  As it is not possible to identify
the criticality of the ``average environment'' by looking at the peak
of a susceptibility, following the seminal paper by Mora and
Bialek\cite{Bialek2011-S} we say that a particular distribution is
critical if it obeys Zipf's law.  To check this, for each
particular environment, we plot the probability of states ordered by
their rank and the energy as a function of the entropy (see
\cite{Bialek2011-S}).  In critical cases, the energy should be a linear
function of the entropy, and the rank ordering should obey the Zipf's
law \cite{Bialek2011-S,Schwab2013-S}.

\begin{figure}[h!]
\centering\includegraphics[width=0.7\textwidth]{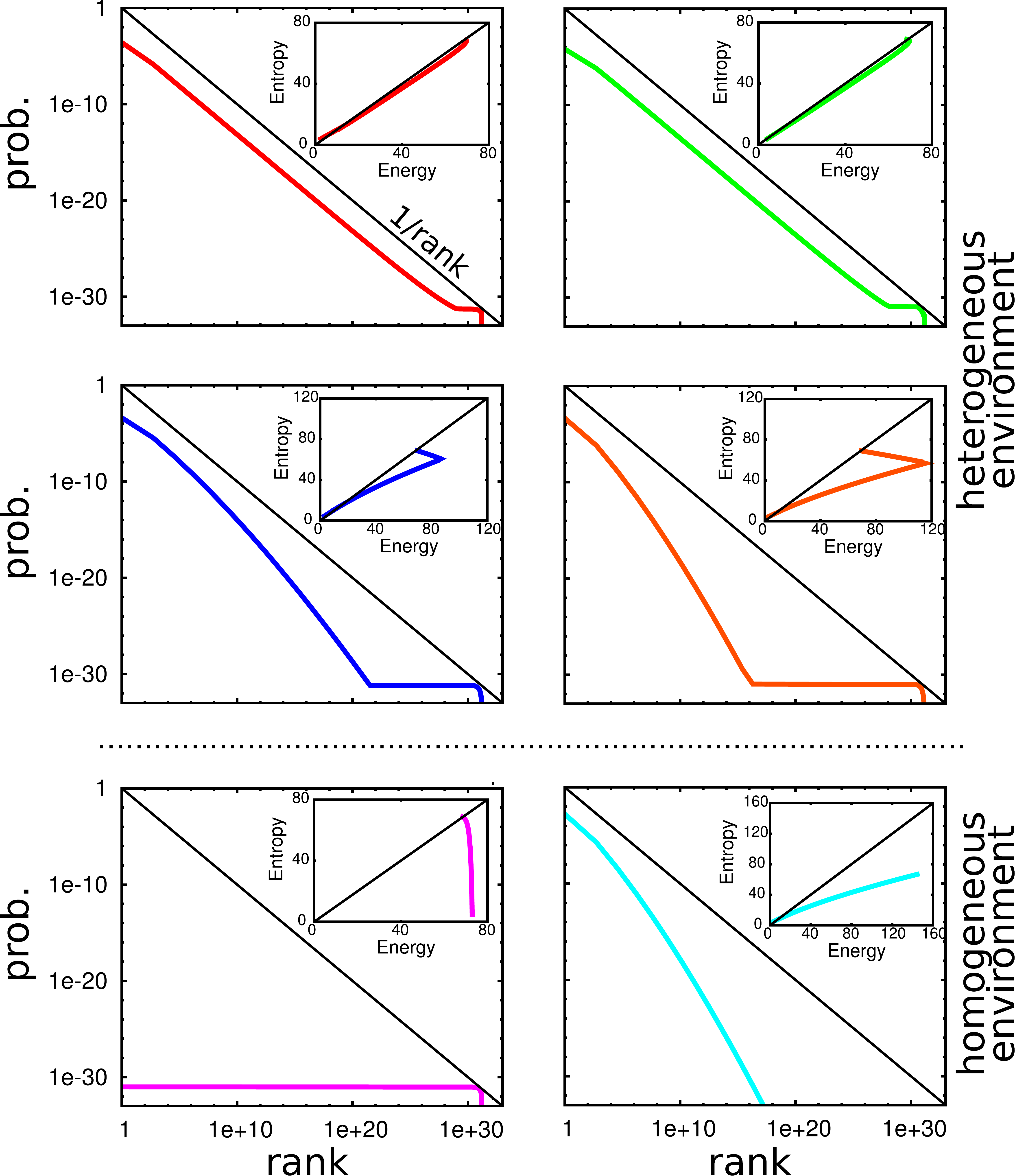}
\caption{ {\it Heterogeneity and criticality of the averaged
    environment.} Each panel of the figure refers to the corresponding
  panel, with the same position and color code, in Fig. 3 of the main
  text. In each panel, the main plot shows the probability of states
  in the ``averaged environment''
  $P_{\mathrm{env}}(\mathbf{s}|\rho_{\mathrm{src}})$ with the states
  $\mathbf{s}$ ranked in order of decreasing probability. The inset
  shows the energy associated to $\bar{P}_{\mathrm{env}}$ as a
  function of the entropy. The black lines define the expected linear
  behavior in the critical case \cite{Bialek2011-S}.  Only the red and
  green settings correspond to a critical distribution obeying also
  Zipf's law ($1/x$). The more interesting cases are the blue and
  the orange ones: for these, the internal distribution is critical
  even though the average environment is not.}
\label{fig:zipf}
\end{figure}

The results are shown in Fig. \ref{fig:zipf}, where we have kept the
same relative position and color code as in the original plot (Fig.3
of the main plot). Only two of the six averaged environments presented
in Fig. 3 of the main text turn out to be critical in the sense of
Zipf's law (upper panels in Fig. \ref{fig:zipf}). Furthermore, in the
two central cases, the averaged environment is not critical, but the
optimal internal distribution peaks around criticality (right panels
in Fig. \ref{fig:zipf}). This case corresponds to an environment
composed, essentially, of two very different type of sources, and
individuals have to accommodate to the critical point to respond to
both of them efficiently.This demonstrates that our
  approach works in a general scenario of heterogeneous sources,
  without requiring the environment to be Zipfian.  Still the
  environment could be critical in some other unspecified (and
  unknown) sense.

 Up to now, we have referred to ``homogeneity'' and
  ``heterogeneity'' of the environment in a rather vague manner.  To
  objectively measure the level of heterogeneity, we can make use of
  the (continuous) Shannon entropy of the distribution of moments
  (observables) in the external distributions.  Given the environment
  characterized by its distribution of sources
  $\rho_{\mathrm{src}}(\boldsymbol{\alpha})$, we compute its
  associated distribution of moments $\varrho_{\mathrm{src}}(\langle
  \boldsymbol{\phi} \rangle)$:
\begin{equation}
\varrho_{\mathrm{src}}(\langle\boldsymbol{\phi}\rangle)=\rho_{\mathrm{src}}(\mathbf{\boldsymbol{\alpha}}) \left|\dfrac{\partial \langle \boldsymbol{\phi} \rangle}{\partial \boldsymbol{\alpha}}
\right|^{-1}.
\end{equation}
Then, the entropy can be computed as
\begin{equation}
 S[\varrho_{\mathrm{src}}] = -\int d\langle\boldsymbol{\phi}\rangle 
\varrho_{\mathrm{src}}(\langle\boldsymbol{\phi}\rangle)\log 
\varrho_{\mathrm{src}}(\langle\boldsymbol{\phi}\rangle).
\end{equation}
Note that, as it is defined in the continuum limit, the entropy can be negative \cite{Cover-Thomas-S}.

\begin{figure}[h!]
\centering\includegraphics[width=0.6\textwidth]{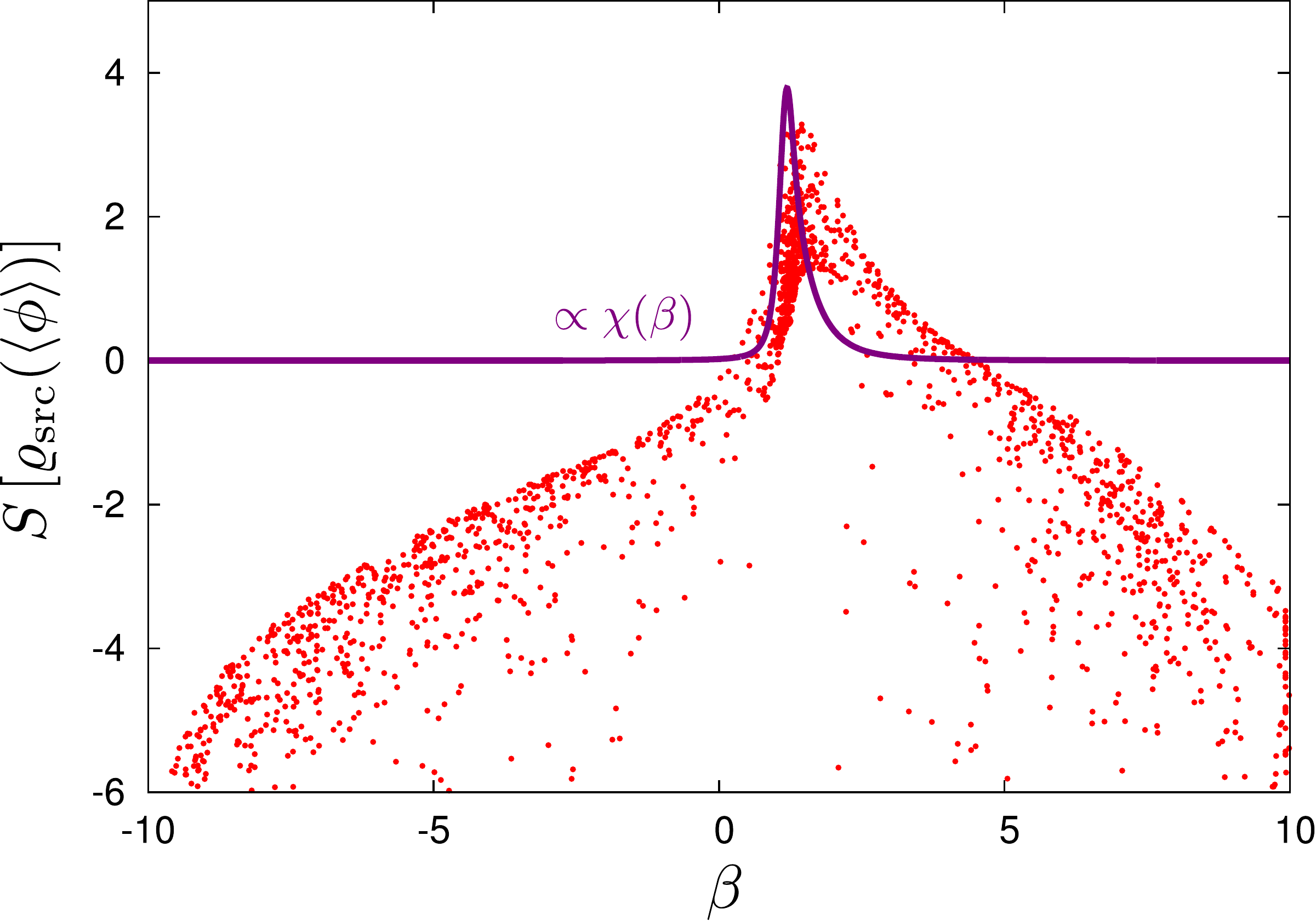}
\caption{{\it Measuring the heterogeneity of the
      environment.} Each dot corresponds to a randomly generated
    distribution of sources $\rho_{\mathrm{src}}(\alpha)$, i.e. an
    environment.  For each of them, we compute the associated optimal
    representation $\beta$ as well as the continuous entropy of its
    moment distribution,
    $S[\varrho_{\mathrm{src}}(\langle\phi\rangle)]$.  Because the space of
    all possible distributions $\rho_{\mathrm{src}}(\alpha)$ can be
    sampled in many different ways, we restrict ourselves to the case of uniform
    distributions with compact support in the range
    $\alpha\in[-10,10]$. The purple
    line is the Fisher Information for the corresponding $\beta$
    (rescaled for visual comparing). We can see that, as the entropy
    increases, the mapping is confined to the critical region, whereas
    it can be anywhere for low entropies. The parameters
    $\alpha,\beta$ refer to the mean-field Ising, eq. \ref{eq:ising},
    with $N=100$.  }
\label{fig:heterogeneity}
\end{figure}

 We have randomly generated different distributions
  $\rho_{\mathrm{src}}(\alpha)$ for the one-parameter case, where the
  parameter $\alpha$ refers to the mean-field Ising model,
  eq. \ref{eq:ising}.  For each distribution
  $\rho_{\mathrm{src}}(\alpha)$, we compute its optimal representation
  $\beta$ as well as the entropy of
  $\varrho_{\mathrm{src}}(\langle\phi\rangle)$. The result is plotted
  in Fig. \ref{fig:heterogeneity} and it is compared with the
  corresponding Fisher Information $\chi(\beta)$.  We see that, when
  the entropy is low (homogeneous environments), the optimal mapping
  $\beta$ can lie at any value, but, when the entropy increases
  (heterogeneous environments), $\beta$ is gradually confined into the
  critical region.  Thus, even though we have not been able to specify 
a quantitative distinction between heterogeneous and homogeneous environments, 
the entropy of the distribution of moments does
constitute a good proxy for the environmental heterogeneity.

\section{Adaptive and Co-adaptive models}
\label{sec:adaptive}

In this section we show that our results are robust in the sense that
criticality is also obtained in a adaptive --rather than
evolutionary-- type of dynamics.

 \subsection{Co-adaptive model}
\begin{figure}[!h]
\centering \includegraphics[width=0.8\textwidth]{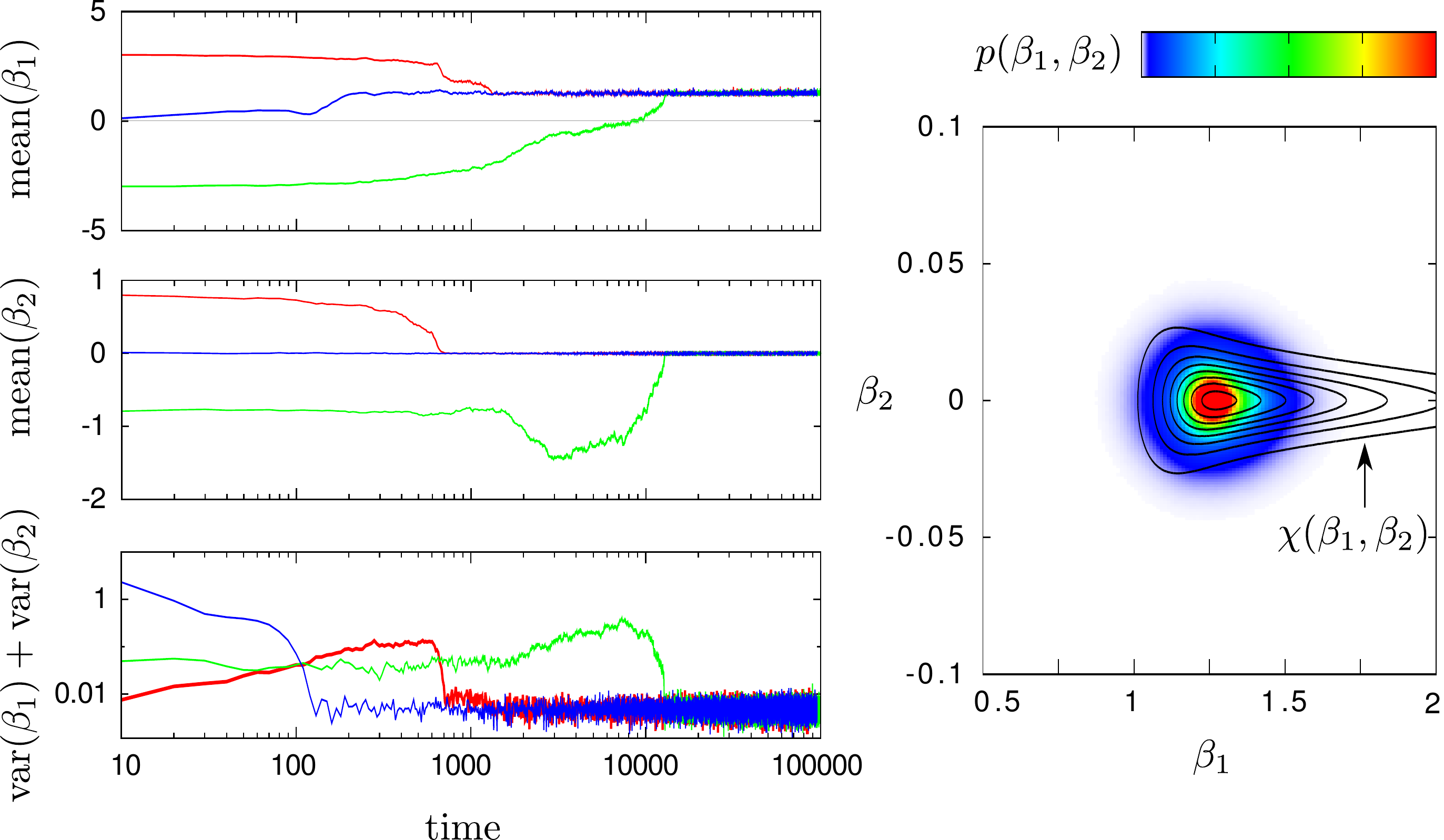}
\caption{ Co-adaptive dynamics. Left panels show the time dependence
  of the first moments of the distribution of internal
  parameters. Independent of initial conditions, the community moves
  toward the critical point ($\beta_1=1$ and $\beta_2=0$) and reaches
  the same stationary solution as the corresponding co-evolutionary
  model. The stationary probability, on the right panel, shows a peak
  where the generalized susceptibility or Fisher information peaks
  (see contour plots).  Parameters, probability distribution
  parametrization, and initial conditions are similar to the ones used
  in Fig. 2 in the main text. $\epsilon_{1}=0.1$, $\epsilon_{2}=0.01$,
  $\nu=0.3$}
\label{fig:co-adaptive}
\end{figure}

We have also developed the following co-adaptive model (counterpart of
the co-evolutionary model):
\begin{enumerate}
\item We start with $M$ individuals with parameters distributed as $p(\boldsymbol{\beta}, t=0)$. Here
	$t$ refers to time.
\item At every time step, every individual can change its parameters
  by a small jump in one of them. Among all the possibilities, it chooses the
  one which minimizes the mean KL divergence to the rest of the
  community:
\begin{equation}
  \boldsymbol{\beta}^i\longrightarrow \argmin_{\boldsymbol{\epsilon} 
    \in E} \dfrac{1}{{M-1}} \sum_{j\neq i}^M D(\boldsymbol{\beta}^j|
  \boldsymbol{\beta}^i+\boldsymbol{\epsilon}), \quad E = \{(\pm
  \epsilon_1,0,. ..0),(0,\pm \epsilon_2,0,...),...\}
\end{equation}
\item We introduce some noise in the adaptation, and with probability
  $\nu$ every individual can change its parameter with a small random
  fluctuation $\xi$, Gaussian distributed with zero mean and deviation
  $\sigma$.
\item Time is incremented as $t\rightarrow t+1$ and the process is iterated.
\end{enumerate}
In Fig. \ref{fig:co-adaptive} we show the evolution of the mean and
variance of $\boldsymbol{\beta}$ in the community with three different
initial conditions (same as in Fig. 2 of the main text), together with
the stationary distribution. Individuals cluster again in the vicinity
of the critical point, and the result is independent of the initial
conditions.

\subsection{Adaptive model}

\begin{figure}[!h]
\centering \includegraphics[width=0.7\textwidth]{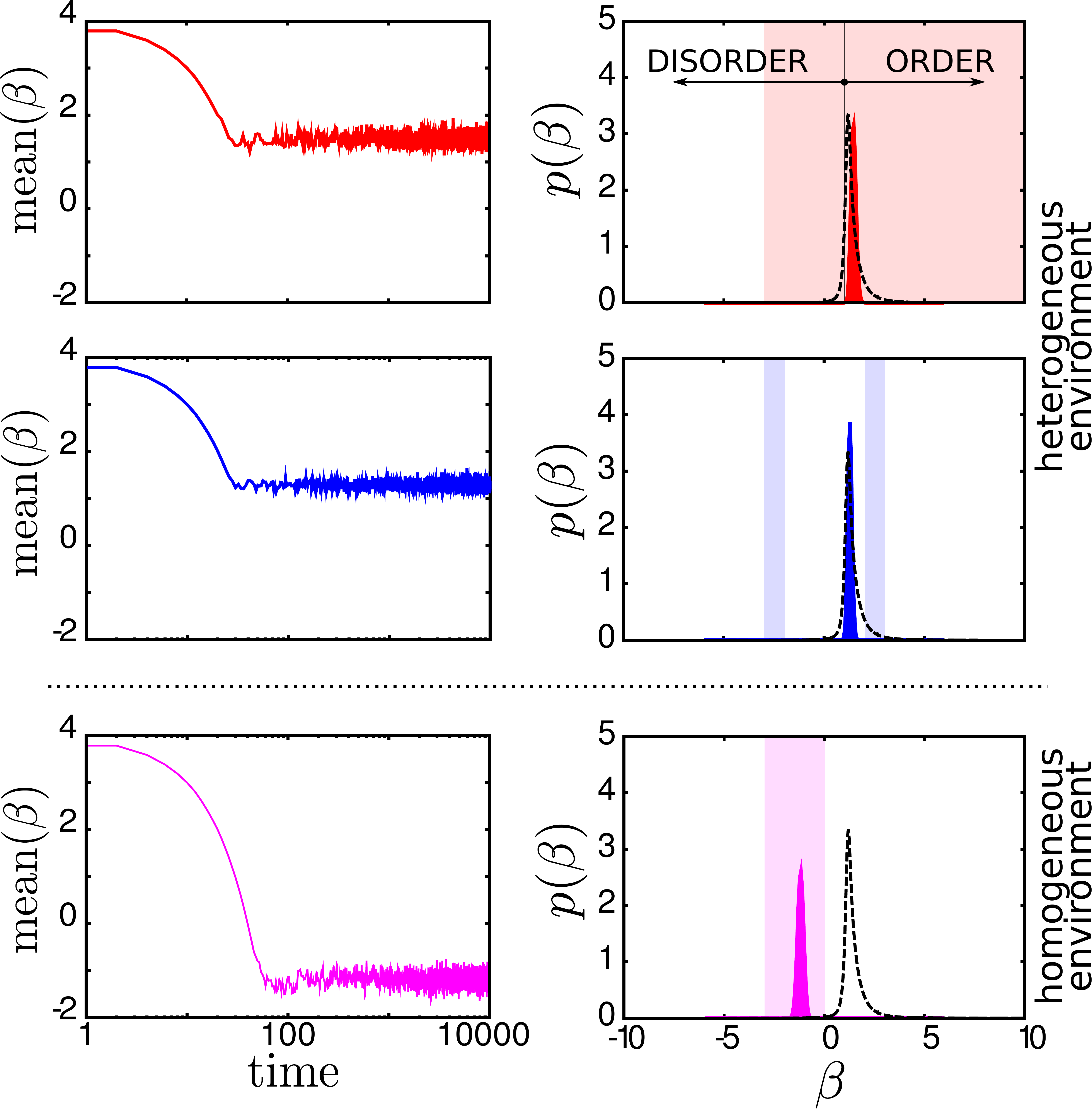}
\caption{Adaptive dynamics in the presence of varying external
  environments.  The figure shows the average of the internal
  parameters in the community and the stationary distributions for
  different environments.  As obtained for the analogous evolutionary
  model, the stationary solution peaks at the critical point if the
  external environment is sufficiently heterogeneous, while it does
  not for simpler environments. Parameters, probability
  parametrization and initial conditions are similar to the ones used
  in Fig. 3 (same color code) of the main text. $\epsilon=0.1$,
  $\nu=0.3$.}
\label{fig:adaptive}
\end{figure}

The adaptive model, in which a agents in a community of $M$
individuals change their parameters to accommodate to an external
environment, proceeds as follows:
\begin{enumerate}
\item As in the Evolutionary models, every source is parametrized by
  eq. \ref{eq:Hsrc} and the individuals by eq. \ref{eq:Hmap}.
\item We start with $M$ individuals whose parameters are distributed
  as $p(\boldsymbol{\beta},t=0)$. Note that, in this model the
  individuals are completely independent, and having a big community
  is only useful in terms of the statistics.
\item At every time step, we generate $S$ external sources from the
  pool $\rho_{\mathrm{src}}(\boldsymbol{\alpha})$.
\item 
At every time step, every individual can change its parameters
  by a small jump in one of them; among all the possibilities it chooses the
  one which minimizes the mean KL divergence to the sources:
  \begin{equation}
  \boldsymbol{\beta}^i\longrightarrow \argmin_{\boldsymbol{\epsilon} 
    \in E} \dfrac{1}{S} \sum_{u=1}^S D(\boldsymbol{\alpha}^u|
  \boldsymbol{\beta}^i+\boldsymbol{\epsilon}),\quad E = \{(\pm
  \epsilon_1,0,. ..0),(0,\pm \epsilon_2,0,...),...\}
\end{equation}
\item With probability $\nu$, we introduce some noise in the
  adaptation, and every individual can change its parameters with a
  Gaussian random fluctuation $\xi$, with mean zero and variance
  $\sigma$.
\item Time is incremented as $t\rightarrow t+1$. Another set of
  sources is generated from $\rho_{\mathrm{src}}(\boldsymbol{\alpha})$
  and the process is iterated.
\end{enumerate}
Fig. \ref{fig:adaptive} shows the dynamics of the community with three
different environments (the same ones as in Fig. 3 of the main
text). As in the analogous evolutionary model, heterogeneous
environments lead to a tuning of parameters to criticality (left and
right panels), while very specific sources (central panels) do not.

\newpage

\clearpage

\section{Supplementary Videos}

{\it Videos are available for download in the src zipped file on the ArXiv.}

\subsection*{Video 1: Simulation of the Co-Evolutionary model leading self-consistently to criticality}

A community of agents or cognitive systems co-evolve according to a genetic algorithm. Different colors represent different initial conditions and so individuals of different colors do not interact
with each other. Each agent constructs a map of the rest of the community, represented with a dot in the two parameter space ($\beta_1$ and $\beta_2$). As in Video 1, individuals with better
representations reproduce more probably, and the offspring inherits the parameters from their parents with small mutations. The simulation shows how, independently of the initial condition, the
individuals self-tune to the maximum of the Fisher information or critical point.

\subsection*{Video 2: Simulation of the Co-Evolutionary model for small systems}

As in Video 1, a community of agents or cognitive systems co-evolve to understand each other. Each agent constructs a map of the environment, represented with a dot in the two parameter space
($\beta_1$ and $\beta_2$). The information is encoded in strings of $N$ binary variables. Every agent has $N=10$ in the upper panel and $N=100$ in the lower one. The Fisher Information for the two
parameters map is plotted in the background.
We can see how, after iterating the genetic algorithm, the agents localize at the maximum of the Fisher Information. When N increases, the peak approaches the critical point.

\subsection*{Video 3: Simulation of the Evolutionary model leading to criticality in complex environments}

A community of agents or cognitive systems in different environments evolve according to a genetic algorithm. Every agent is represented with a black dot on the vertical axis.
At each time step, different sources are generated from the colored region, every one characterized by a parameter $\beta$. The agents construct an internal map of the sources, represented in their
own internal parameter $\beta$. Individuals with better representations have more chances to reproduce, and the offspring inherits the parameter $\beta$ from their parents with a small mutation.
We can see that, when the source pools are heterogeneous, as occurs for the left and right panels, the community evolves near the maximum of the Fisher Information, or, in other words, the critical
point. However, when the sources are very specific, the agents do not become critical, as occur for the central panels.

\end{document}